\documentclass[12pt]{article}
\usepackage{graphicx}
\usepackage{amsmath}
\usepackage{amsfonts}
\usepackage{amssymb}

\begin{document}

\title{Calculations for Extended Thermodynamics of dense gases up to whatever order and with
all the symmetries}

\author{M.C. Carrisi, S. Pennisi \\
Dipartimento di Matematica ed Informatica , Universit\'{a} di Cagliari, Cagliari,
Italy\\
spennisi@unica.it;cristina.carrisi@tiscali.it}
\date{}
\maketitle \vspace{0.5 cm}
 \small {\em \noindent } \\

\begin{abstract}
The 14 moments model for dense gases, introduced in the last years by Arima,  Taniguchi
Ruggeri, Sugiyama, is here considered. They have found the closure of the balance
equations up to second order with respect to equilibrium; here the closure is found up to
whatever order with respect to equilibrium, but for a more constrained system where more
symmetry conditions are imposed and this in agreement with the suggestion of the kinetic
theory. The results, when restricted at second order with respect to equilibrium, are the
same of the previously cited model but under the further restriction of full symmetries.
\end{abstract}
\textbf{AMS Subject Classification:} \\
\textbf{Key Words:}

\section{Introduction}

Starting point of this research is the article  \cite{1} which belongs to the framework
of Extended Thermodynamics. Some of the original papers on this subject are \cite{2},
\cite{3} while more recent papers are \cite{4}-\cite{18} and the theory has the advantage
to furnish hyperbolic field equations, with finite
speeds of propagation of shock waves and very interesting analytical  properties. \\
It starts from a given set of balance equations where some arbitrary functions appear;
restrictions on these arbitrariness are obtained by imposing the entropy principle and
the relativity principle. \\
However, these restrictions were so strong to allow only particular state functions; for
example, the function $p=p( \rho , T)$ relating the pressure $p$ with the mass density
$\rho$ and the absolute temperature $T$, was determined except for a single variable
function so that it was adapt to describe only particular gases or continuum. \\
This drawback has been overcome in \cite{1} and other articles such as
\cite{19}-\cite{34} by considering two blocks of balance equations; for example, in the
14 moments case treated in \cite{1}, they are
\begin{eqnarray}\label{2.1}
&{}&  \quad \quad \quad \, \, \partial_t F^N + \partial_k F^{kN} = P^N \quad , \quad
\partial_t G^E +
\partial_k G^{kE}
= Q^E \, ,  \\
&{}& \nonumber \\
 &{}&  \mbox{where} \quad F^N = (F, F^i, F^{ij}) \quad \, \quad , \quad G^E = (G,
G^i) \quad ,
\nonumber \\
&{}&  \quad \quad \quad \, \, F^{kN} = (F^k, F^{ki}, F^{kij}) \quad , \quad G^{kE} =
(G^k,
G^{ki}) \quad , \nonumber \\
&{}&  \quad \quad \quad  \, \, P^{N} = (0, 0, P^{ij}) \quad\quad\quad , \quad Q^{E} = (0,
Q^{i}) \quad . \nonumber
\end{eqnarray}
The first 2 components of $P^N$ are zero because the first 2 components of equations
$(\ref{2.1})_1$ are the conservation laws of mass and momentum; the first  component of
$Q^E$ is zero because the first  component of equations $(\ref{2.1})_2$ is the
conservation laws of energy. The whole block $(\ref{2.1})_2$ can be considered an "Energy
Block". \\
The equations $(\ref{2.1})$ can be written in a more compact form as
\begin{eqnarray}\label{2.2}
&{}&  \quad \quad \quad \, \, \partial_t F^A + \partial_k F^{kA} = P^A \quad ,
 \\
&{}& \nonumber \\
 &{}&  \mbox{where} \quad F^A = (F^N, G^E) \quad  , \quad F^{kA} = (F^{kN}, G^{kE}) \quad , \quad
 P^{A} = (P^N, Q^E) \quad . \nonumber
\end{eqnarray}
In the whole set (\ref{2.2}), $F^A$ are the independent variables, while $F^{kij}$,
$G^{ki}$, $P^{ij}$, $Q^{i}$ are constitutive functions. Restrictions on their
generalities are obtained by imposing
\begin{enumerate}
  \item \textbf{The Entropy Principle} which guarantees the existence of an entropy density $h$
  and an entropy flux $h^k$ such that the equation
  \begin{eqnarray}\label{2.3}
&{}&   \partial_t h + \partial_k h^{k} = \sigma \geq 0  \quad ,
\end{eqnarray}
holds whatever solution of the equations (\ref{2.2}). \\
Thanks to Liu' s Theorem \cite{35}, \cite{36}, this is equivalent to assuming the
existence of Lagrange Multipliers $\mu_A$ such that
  \begin{eqnarray}\label{2bis.1}
&{}&   d \, h= \mu_A d \, F^A  \quad , \quad d \, h^k= \mu_A d \, F^{kA} \quad , \quad
\sigma = \mu_A P^A \, .
\end{eqnarray}
An idea conceived by Ruggeri is to define the 4-potentials $h'$, $h'^k$ as
  \begin{eqnarray}\label{2bis.2}
&{}&  h'= \mu_A  F^A - h \quad , \quad h'^k= \mu_A F^{kA} - h^k \quad ,
\end{eqnarray}
so that eqs. $(\ref{2bis.1})_{1,2}$ become
  \begin{eqnarray*}
&{}&  d \, h'= F^A d \, \mu_A   \quad , \quad d \, h'^k= F^{kA}  d \, \mu_A \quad ,
\end{eqnarray*}
which are equivalent to
  \begin{eqnarray}\label{2bis.3}
&{}&   F^A  = \frac{\partial h'}{\partial  \mu_A } \quad , \quad F^{kA} = \frac{\partial
h'^k}{\partial  \mu_A } \quad ,
\end{eqnarray}
if the Lagrange Multipliers are taken as independent variables. A nice consequence of
eqs. $(\ref{2bis.3})$ is that the field equations assume the symmetric form. \\
Other restrictions are given by
\item \textbf{The symmetry conditions}, that is the second component of $F^N$ is equal to the
first component of $F^{kN}$, the third component of $F^N$ is equal to the second
component of $F^{kN}$, the second component of $G^E$ is equal to the first component of
$G^{kE}$. Moreover, $F^{ij}$, $F^{kij}$ and $G^{ki}$ are symmetric tensors.  The symmetry
of  $F^{kij}$ and $G^{ki}$ is motivated by the kinetic counterpart of this theory (see
section 4 of ref. \cite{19}), even if  was not imposed in \cite{1} in order to have a
more general model. We propose, in a future article, to remove this further constraint.
Thanks to eqs. (\ref{2bis.3}) these conditions may be expressed as
\begin{eqnarray}\label{3.0}
&{}&  \frac{\partial h'}{\partial  \mu_i } =  \frac{\partial h'^i}{\partial  \mu } \quad
, \quad \frac{\partial h'}{\partial  \mu_{ij} } =  \frac{\partial h'^i}{\partial  \mu_j }
\quad , \quad \frac{\partial h'}{\partial  \lambda_i } =  \frac{\partial h'^i}{\partial
\lambda } \quad , \quad  \frac{\partial h'^{[k}}{\partial  \mu_{i]j} } = 0 \quad ,
\frac{\partial h'^{[k}}{\partial  \lambda_{i]} } = 0 \quad ,
\end{eqnarray}
where we have assumed the decomposition $\mu_A= ( \mu , \mu_i , \mu_{ij} , \lambda ,
\lambda_i )$ for the Lagrange Multipliers. Moreover $\mu_{ij}$ is a symmetric tensor. \\
The next conditions come from
  \item \textbf{The Galilean Relativity Principle}.
\end{enumerate} We prefer to devote an entire section, the next one, to describe how to
impose this principle. The result, combined with the above conditions  coming from the
Entropy Principle and the Symmetry Conditions will be that a scalar function $H$ exists,
such that
\begin{eqnarray}\label{8.2}
&{}&   h' = \frac{\partial H}{\partial  \mu } \quad , \quad h'^i = \frac{\partial
H}{\partial  \mu_i } \quad .
\end{eqnarray}
\begin{eqnarray}\label{9.1}
&{}&  \frac{\partial^2 H}{\partial  \mu \partial  \mu_{ij}} = \frac{\partial^2
H}{\partial \mu_i \partial  \mu_{j}} \quad , \quad \frac{\partial^2 H}{\partial  \mu
\partial  \lambda_i} = \frac{\partial^2 H}{\partial \lambda \partial  \mu_{i}} \quad , \quad
\frac{\partial^2 H}{\partial  \mu_{[k} \partial \mu_{i]j} } = 0 \quad , \quad
\frac{\partial^2 H}{\partial  \mu_{[k} \partial \lambda_{i]} } = 0\, .
\end{eqnarray}
\begin{eqnarray}\label{9.3}
&{}& \frac{\partial^2 H}{\partial  \mu \partial \mu_k } \mu_i + 2 \frac{\partial^2
H}{\partial \mu
\partial  \mu_{kj}}\mu_{ji}+ 2 \frac{\partial^2
H}{\partial \mu
\partial  \mu_{ki}}\lambda +
2 \frac{\partial^2 H}{\partial  \mu_k \partial  \mu_{ij}} \lambda_{j} + \frac{\partial^2
H}{\partial  \mu \partial  \lambda_k} \lambda_i + \frac{\partial H}{\partial  \mu}
\delta^{ki}=0 \, .
\end{eqnarray}
In section 3 of the present article, we find the general solution up to whatever order
with respect to equilibrium, of the conditions (\ref{9.1}), (\ref{9.3}). \\
In section 4 we will see the implications of this solution to a second order theory,
coming back to the moments as independent variables; this will allow us to recover, as
first order approximation for the functions $F^{kij}$ and $G^{ki}$, the same result of
\cite{1}; similarly, for the second order approximation for the entropy density and its
flux. So the present results generalize that article and also confirm it, because we
guarantee that the equations at subsequent orders don' t give other restrictions on the
first order theory.

\section{The Galilean Relativity Principle}
  There are two ways to impose this principle. One of these is to decompose the variables
  $F^{A}$, $F^{kA}$, $P^{A}$, $\mu_{A}$ in their corresponding non convective parts $\hat{F}^{A}$,
  $\hat{F}^{kA}$, $\hat{P}^{A}$, $\hat{\mu}_{A}$ and in velocity dependent parts, where
  the velocity is defined by
    \begin{eqnarray}\label{3.1}
&{}&  v^i = F ^{-1}F^i \quad .
\end{eqnarray}
This decomposition can be written as
\begin{eqnarray}\label{3.2}
&{}& F^{A} = {X^A}_B( \vec{v}) \hat{F}^{B} \quad , \quad F^{kA} - v^k F^A = {X^A}_B(
\vec{v}) \hat{F}^{kB} \quad , \quad P^{A} = {X^A}_B( \vec{v}) \hat{P}^{B} \quad , \\
&{}& h' = \hat{h}' \quad , \quad h'^{k} - v^k h' =  \hat{h}'^{k} \quad , \quad
\hat{\mu}_{A} = \mu_B {X^B}_A( \vec{v})  \quad , \nonumber
\end{eqnarray}
where
\begin{eqnarray}\label{3.3}
&{}& {X^A}_B( \vec{v}) =
\begin{pmatrix}
  1 & 0 & 0 & 0 & 0 \\
  v^i & \delta^{i}_a & 0 & 0 & 0 \\
  v^i v^j & 2v^{(i} \delta^{j)}_a & \delta^i_{(a}\delta^j_{b)} & 0 & 0 \\
  v^2 & 2 v_a & 0 & 1 & 0 \\
  v^2 v^i & v^2 \delta^i_a + 2 v^i v_a & 2 \delta^i_{(a} v_{b)} & v^i & \delta^i_a
\end{pmatrix}
\end{eqnarray}
After that, all the conditions are expressed in terms of the non convective parts of the
variables. \\
This procedure is described in \cite{2}, \cite{36} for the case considering only the
block $(\ref{2.1})_1$ and is followed in \cite{1} for the whole set (\ref{2.1}). \\
Another way to impose this principle leads to easier calculations; it is described in
\cite{37} for the case considering only the the block
$(\ref{2.1})_1$ and here we show how it is adapt also for the whole set (\ref{2.1}). \\
First of all, we need to know the transformation law of the variables between two
reference frames moving one with respect to the other with a translational motion with
constant translational velocity $ \vec{v}_\tau$. To know it, we may rewrite (\ref{3.2})
in both frames, that is
\begin{eqnarray}\label{4.1}
&{}& F^{A}_a = {X^A}_B( \vec{v}_a) \hat{F}^{B} \quad , \quad F^{kA}_a - v^k_a F^A_a =
{X^A}_B(
\vec{v}_a) \hat{F}^{kB} \quad ,  \\
&{}& h'_a = \hat{h}' \quad , \quad h'^{k}_a - v^k_a h'_a =  \hat{h}'^{k} \quad , \quad
\hat{\mu}_{A}^a = \mu_B^a {X^B}_A( \vec{v}_a)  \quad , \nonumber
\end{eqnarray}
\begin{eqnarray*}
&{}& F^{A}_r = {X^A}_B( \vec{v}_r) \hat{F}^{B} \quad , \quad F^{kA}_r - v^k_r F^A_r =
{X^A}_B(
\vec{v}_r) \hat{F}^{kB} \quad ,  \\
&{}& h'_r = \hat{h}' \quad , \quad h'^{k}_r - v^k_r h'_r =  \hat{h}'^{k} \quad , \quad
\hat{\mu}_{A}^r = \mu_B^r {X^B}_A( \vec{v}_r)  \quad ,
\end{eqnarray*}
where the index $a$ denotes quantities in the absolute reference frame and index $r$
denotes quantities in the relative one; $\hat{F}^{B}$,
  $\hat{F}^{kB}$, $\hat{h}'$, $\hat{h}'^{k}$,  $\hat{\mu}_{B}$ haven't the index $a$, nor
  the index $r$ because they are independent from the reference frame. \\
  Now we can use a property of the matrix ${X^A}_B( \vec{v})$ which is a consequence of
  its definition (\ref{3.3}) and reads
\begin{eqnarray}\label{4.2}
&{}& {X^C}_A(- \vec{v}) {X^A}_B( \vec{v}) = \delta^C_B \quad .
\end{eqnarray}
So we may contract $(\ref{4.1})_{6,7}$ with ${X^C}_A(- \vec{v}_r)$ so obtaining
\begin{eqnarray*}
&{}& \hat{F}^{C} = {X^C}_A(- \vec{v}_r) F^A_r  \quad , \quad \hat{F}^{kC} = {X^C}_A(-
\vec{v}_r) (F^{kA}_r - v^k_r F^A_r)
\end{eqnarray*}
which can be substituted in $(\ref{4.1})_{1,2}$. The result is
\begin{eqnarray}\label{4.3}
&{}& F^A_a = {X^A}_B( \vec{v}_a) {X^B}_C(- \vec{v}_r) F^C_r  \quad , \quad F^{kA}_a -
v^k_a F^A_a = {X^A}_B( \vec{v}_a) {X^B}_C(- \vec{v}_r) (F^{kC}_r - v^k_r F^C_r) \, .
\quad \quad \quad
\end{eqnarray}
Now we use another property of the matrix ${X^A}_B( \vec{v})$ which is a consequence of
  its definition (\ref{3.3}) and reads
\begin{eqnarray}\label{5.1}
&{}& {X^A}_B(\vec{u}) {X^B}_C( \vec{w}) = {X^A}_C( \vec{u} + \vec{w}) \quad .
\end{eqnarray}
Moreover, we use the well known property
\begin{eqnarray}\label{5.2}
&{}& \vec{v}_a = \vec{v}_r + \vec{v}_\tau  \quad .
\end{eqnarray}
In this way the equations (\ref{4.3}) become
\begin{eqnarray}\label{5.2a}
&{}& F^A_a = {X^A}_C( \vec{v}_\tau)  F^C_r  \quad , \\
&{}& F^{kA}_a - v^k_a F^A_a = {X^A}_C( \vec{v}_\tau)  F^{kC}_r - v^k_r {X^A}_C(
\vec{v}_\tau)  F^C_r \, , \quad \quad \quad \nonumber
\end{eqnarray}
In eq. $(\ref{5.2a})_2$ we can substitute ${X^A}_C( \vec{v}_\tau)  F^C_r$ from eq.
$(\ref{5.2a})_1$ so that it becomes
\begin{eqnarray}\label{5.3}
&{}& F^{kA}_a - v^k_\tau F^A_a = {X^A}_C( \vec{v}_\tau)  F^{kC}_r \, . \quad \quad \quad
\end{eqnarray}
Finally, we deduce $\hat{h}'$, $\hat{h}'^{k}$ and $\hat{\mu}_A$ from
$(\ref{4.1})_{8,9,10}$ and substitute them in $(\ref{4.1})_{3,4,5}$ so obtaining
\begin{eqnarray}\label{5.4}
&{}& h'_a = h'_r \quad , \quad h'^{k}_a - v^k_\tau h' =  h'^{k}_r \quad , \quad \mu^r_{C}
= \mu_B^a {X^B}_C( \vec{v}_\tau)  \quad ,
\end{eqnarray}
where for the last one we have also used a contraction with ${X^A}_C(- \vec{v}_r)$. \\
Well, eqs. $(\ref{5.2a})_1$,  $(\ref{5.3})$ and $(\ref{5.4})$ give the requested
transformation law between the two reference frames and it is very interesting that it
looks like eqs. $(\ref{3.2})$. \\
Now, if the Lagrange Multipliers are taken as independent variables, eqs. $(\ref{5.4})_3$
are only a change of independent variables from $\mu^a_B$ to $\mu^r_C$, while
$(\ref{5.2a})_1$, $(\ref{5.3})$, $(\ref{5.4})_{1,2}$ are conditions because they involve
constitutive functions
\begin{eqnarray}\label{6.1}
&{}& F^A_a = F^A(\mu^a_B) \, , \, F^{kA}_a = F^{kA}(\mu^a_B) \, , \, h'_a  = h'(\mu^a_B)
\, , \, h'^k_a  = h'^k(\mu^a_B) \, , \\
&{}& F^A_r = F^A(\mu^r_B) \, , \, F^{kA}_r = F^{kA}(\mu^r_B) \, , \, h'_r  = h'(\mu^r_B)
\, , \, h'^k_r  = h'^k(\mu^a_B) \, , \nonumber
\end{eqnarray}
where the form of the functions $F^{A}$, $F^{kA}$, $ h'$, $ h'^k$ don' t depend on the
reference frame for the Galilean Relativity Principle. If we substitute $\mu^a_B$ from
eq. $(\ref{5.4})_3$ in $(\ref{6.1})_{1-4}$ and then substitute the result in
$(\ref{5.2a})_1$, $(\ref{5.3})$, $(\ref{5.4})_{1,2}$, we obtain
\begin{eqnarray}\label{6.2}
&{}& F^A(\mu_C^r {X^C}_B(- \vec{v}_\tau)) = {X^A}_C( \vec{v}_\tau) F^C_r \, , \\
&{}& F^{kA}(\mu_C^r {X^C}_B(- \vec{v}_\tau)) - v^k_\tau {X^A}_C( \vec{v}_\tau) F^C_r =
{X^A}_C( \vec{v}_\tau) F^{kC}_r \, , \nonumber \\
&{}& h'(\mu_C^r {X^C}_B(- \vec{v}_\tau)) = h'_r \, , \nonumber \\
&{}& h'^k(\mu_C^r {X^C}_B(- \vec{v}_\tau)) - h' v^k_\tau = h'^k_r \, . \nonumber
\end{eqnarray}
Well, these expressions calculated in $v^i_\tau =0$ are nothing more than eqs.
$(\ref{6.1})_{5-8}$, as we expected. But, for the Galilean Relativity Principle they must
be coincident for whatever value of $v^i_\tau$; this amounts to say that the derivatives
of (\ref{6.2}) with respect to $v^i_\tau$ must hold. \\
This constraint can be written explicitly more easily if we take into account that
$\mu_C^r {X^C}_B(- \vec{v}_\tau)= \mu^a_B$ which can be written explicitly by use of
(\ref{3.3}) and reads
\begin{eqnarray}\label{7.1}
&{}& \mu^a= \mu^r - \mu_i^r v^i_\tau + \mu_{ij}^r v^i_\tau v^j_\tau + \lambda^r v^2_\tau
- \lambda_i^r  v^i_\tau v^2_\tau \, , \\
&{}& \mu^a_h= \mu^r_h - 2 \mu_{ih}^r v^i_\tau  -2  \lambda^r v_{\tau h} +
\lambda_i^r ( v^2_\tau \delta_{h}^i + 2 v^i_\tau v_{\tau h} ) \, , \nonumber\\
&{}& \mu^a_{hk}= \mu^r_{hk} - 2 \lambda_i^r v_{\tau (h} \delta_{k)}^i  \, , \nonumber\\
&{}& \lambda^a = \lambda^r - \lambda_i^r v^i_{\tau} \, , \nonumber \\
&{}& \lambda^a_h = \lambda^r_h \, , \nonumber
\end{eqnarray}
from which
\begin{eqnarray}\label{7.2}
&{}& \frac{\partial \mu^a}{\partial v^i_{\tau}}= - \mu^a_i \quad , \quad \frac{\partial
\mu^a_h}{\partial v^i_{\tau}}= - 2 \mu_{ih}^a   -2  \lambda^a \delta_{hi}  \quad , \\
&{}& \frac{\partial\mu^a_{hk}}{\partial v^i_{\tau}}= - 2 \lambda_{(h}^a \delta_{k)i}
\quad , \quad \frac{\partial \lambda^a}{\partial v^i_{\tau}} =  - \lambda_i^a \quad ,
\quad \frac{\partial \lambda^a_h}{\partial v^i_{\tau}} = 0 \, . \nonumber
\end{eqnarray}
Consequently, the derivatives of $(\ref{6.2})_{3,4}$ with respect to $ v^i_{\tau}$ become
\begin{eqnarray}\label{7.3}
&{}& \frac{\partial h'}{\partial \mu} \mu_i + \frac{\partial h'}{\partial \mu_h}(2
\mu_{ih} +2  \lambda \delta_{hi})  + 2 \frac{\partial h'}{\partial \mu_{hi}} \lambda_{h}
+ \frac{\partial h'}{\partial \lambda} \lambda_i=0 \, , \\
&{}& \frac{\partial h'^k}{\partial \mu} \mu_i + \frac{\partial h'^k}{\partial \mu_h}(2
\mu_{ih} +2  \lambda \delta_{hi})  + 2 \frac{\partial h'^k}{\partial \mu_{hi}}
\lambda_{h} + \frac{\partial h'^k}{\partial \lambda} \lambda_i + h' \delta^{ki}=0 \, ,
\nonumber
\end{eqnarray}
where we have omitted the index $a$ denoting variables in the absolute reference frame
because they remain unchanged if we change $ v^i_{\tau}$ with $- v^i_{\tau}$, that is, if
we exchange the absolute and the relative reference frames. \\
It is not necessary to impose the derivatives of $(\ref{6.2})_{1,2}$ with respect to $
v^i_{\tau}$  because they are consequences of (\ref{7.3}) and (\ref{2bis.3}). \\
Consequently, the Galilean Relativity Principle amounts simply in the 2 equations
(\ref{7.3}). \\
So we have to find the most general functions satisfying (\ref{3.0}) and (\ref{7.3}).
After that, we have to use eqs. $(\ref{2bis.3})_1$ to obtain the Lagrange Multipliers in
terms of the variables $F^A$. By substituting them in $(\ref{2bis.3})_2$ and in $h'=h'(
\mu_A)$, $h'^k=h'^k( \mu_A)$ we obtain the constitutive functions in terms of the
variables $F^A$. If we want the non convective parts of our expressions, it suffices to
calculate the left hand side of eqs. $(\ref{2bis.3})_1$ in $\vec{v}=\vec{0}$ so that they
become
\begin{eqnarray}\label{8.1}
&{}&   \hat{F}^A  = \frac{\partial h'}{\partial  \mu_A } \quad .
\end{eqnarray}
From this equation we obtain the Lagrange Multipliers in terms of $\hat{F}^A$ (Obviously,
they will be $\hat{\mu}_A$) and after that substitute them in $h'=h'( \mu_A)$,
$h'^k=h'^k( \mu_A)$ (the last of which will in effect be $\hat{h}'^k$) and into
$\hat{F}^{kA}  = \frac{\partial h'^k}{\partial  \mu_A }$, that is eq. $(\ref{2bis.3})_2$
calculated in $\vec{v}=\vec{0}$. \\
It has to be noted that from (\ref{3.1}) it follows $\hat{F}^i=0$, so that one of the
equations (\ref{8.1}) is $ 0 = \frac{\partial h'}{\partial  \mu_i }$; this doesn' t mean
that $h'$ doesn' t depend on $\mu_i$, but this is simply an implicit function defining
jointly with the other equations (\ref{8.1}) the quantities $\hat{\mu}_A$ in terms of
$\hat{F}^A$. We note also here the ground to settle $\mu_i=0$ at equilibrium: in fact, in
this state we have $\mu_{ij}=0$, $\lambda_{i}=0$ so that, for the Representation
Theorems, $\frac{\partial h'}{\partial  \mu_i }$ is proportional to $\mu_i$ and
$\frac{\partial h'}{\partial  \mu_i }=0$ implies $\mu_i=0$. \\
By using a procedure similar to that of the paper \cite{37}, we can prove that we obtain
the same results of the firstly described approach. \\
Now, from $(\ref{3.0})_2$ it follows $\frac{\partial h'^{[i}}{\partial  \mu_{j]} }=0$;
this equation, together with $(\ref{3.0})_1$ are equivalent to assuming the existence of
a scalar function $H$ such that the above mentioned (\ref{8.2}) holds. In fact, the
integrability conditions for (\ref{8.2}) are exactly $(\ref{3.0})_1$ and
$\frac{\partial h'^{[i}}{\partial  \mu_{j]} }=0$. \\
Thanks to (\ref{8.2}), we can rewrite (\ref{3.0}) and (\ref{7.3}) as the above mentioned
eq. (\ref{9.1}) and
\begin{eqnarray}\label{9.2}
&{}& \frac{\partial^2 H}{\partial  \mu^2 } \mu_i + \frac{\partial^2 H}{\partial  \mu
\partial  \mu_{h}}(2 \mu_{ih} +2  \lambda \delta_{hi})  +
2 \frac{\partial^2 H}{\partial  \mu \partial  \mu_{hi}} \lambda_{h}
+ \frac{\partial^2 H}{\partial  \mu \partial  \lambda} \lambda_i=0 \, , \\
&{}& \frac{\partial^2 H}{\partial  \mu \partial \mu_k } \mu_i + \frac{\partial^2
H}{\partial \mu_h
\partial  \mu_{k}}(2 \mu_{ih} +2  \lambda \delta_{hi})  +
2 \frac{\partial^2 H}{\partial  \mu_k \partial  \mu_{hi}} \lambda_{h} + \frac{\partial^2
H}{\partial  \mu_k \partial  \lambda} \lambda_i + \frac{\partial H}{\partial  \mu}
\delta^{ki}=0 \, . \nonumber
\end{eqnarray}
We note now that the derivative of $(\ref{9.2})_1$ with respect to $\mu_k$ is equal to
the derivative of $(\ref{9.2})_2$ with respect to $\mu$; similarly, the derivative of
$(\ref{9.2})_1$ with respect to $\lambda_k$ is equal to the derivative of $(\ref{9.2})_2$
with respect to $\lambda$, as it can be seen by using also eqs. (\ref{9.1}). \\
Consequently, the left hand side of eq. $(\ref{9.2})_1$ is a vectorial function depending
only on two scalars $\mu$, $\lambda$ and on a symmetric tensor $\mu_{ij}$. For the
Representation Theorems \cite{38}-\cite{46}, it can be only zero. \\
In other words, eq. $(\ref{9.2})_1$ is a consequence of $(\ref{9.1})$ and
$(\ref{9.2})_2$, so that it has not to be imposed. By using eqs. (\ref{9.1}) we can
rewrite eq. $(\ref{9.2})_2$ as the above mentioned (\ref{9.3}).

\section{The solution up to whatever order}
Let us firstly show a particular solution of (\ref{9.1}) and (\ref{9.3}). \\
Let $\psi_n(\mu , \lambda)$ be a family of functions constrained by
\begin{eqnarray}\label{11.5}
\frac{\partial}{\partial  \mu } \psi_{n+1} = \psi_n \quad \mbox{for} \quad n \geq 0 \, .
\end{eqnarray}
Let us define the function
\begin{eqnarray}\label{11.6}
H_1= \sum_{p,q}^{0 \cdots \infty} \sum_{r \in I_{p}} \frac{1}{p!} \frac{1}{q!}
\frac{1}{r!} \frac{(p+2q+r+1)!!}{p+2q+r+1} \frac{\partial^{r+p}}{\partial \lambda^r
\partial \mu^p} \left[ \left( \frac{-1}{2 \lambda} \right)^{q+\frac{p+r}{2}}
\psi_{\frac{p+r}{2}}\right]
\cdot \\
\cdot \delta^{(i_1 \cdots i_{p}h_1k_1 \cdots h_qk_qj_1 \cdots j_r)} \mu_{i_1} \cdots
\mu_{i_{p}}\mu_{h_1k_1} \cdots \mu_{h_qk_q} \lambda_{j_1} \cdots \lambda_{j_r}  \, .
\nonumber
\end{eqnarray}
In Appendix 1 we will show that eqs. (\ref{9.1}) and (\ref{9.3})  are satisfied if $H$ is
replaced by $H_1$; in other words, $H=H_1$ is a particular solution of our conditions. \\
Moreover, $(H_1)_{eq.}= \psi_0(\mu , \lambda)$ which is an arbitrary two-variables
function, such as $H_{eq.}$. So we can identify
\begin{eqnarray}\label{11.7}
\psi_0(\mu , \lambda)=H_{eq.}
\end{eqnarray}
and define
\begin{eqnarray}\label{11.8}
\Delta H =H- H_1\, .
\end{eqnarray}
In this way the conditions (\ref{9.1}) and (\ref{9.3}) become
\begin{eqnarray}\label{11.9}
&{}&  \frac{\partial^2 \Delta H}{\partial  \mu \partial  \mu_{ij}} = \frac{\partial^2
\Delta H}{\partial \mu_i \partial  \mu_{j}} \quad , \quad \frac{\partial^2 \Delta
H}{\partial \mu
\partial  \lambda_i} = \frac{\partial^2 \Delta H}{\partial \lambda \partial  \mu_{i}} \quad
,  \quad \frac{\partial^2  \Delta H}{\partial  \mu_{[k} \partial \mu_{i]j} } = 0 \quad ,
\quad
\frac{\partial^2  \Delta H}{\partial  \mu_{[k} \partial \lambda_{i]} } = 0\, .\\
&{}& \frac{\partial^2 \Delta H}{\partial  \mu \partial \mu_k } \mu_i + 2 \frac{\partial^2
\Delta H}{\partial \mu
\partial  \mu_{kj}}\mu_{ji}+ 2 \frac{\partial^2
\Delta H}{\partial \mu
\partial  \mu_{ki}}\lambda +
2 \frac{\partial^2 \Delta H}{\partial  \mu_k \partial  \mu_{ij}} \lambda_{j} +
\frac{\partial^2 \Delta H}{\partial  \mu \partial  \lambda_k} \lambda_i + \frac{\partial
\Delta H}{\partial  \mu} \delta^{ki}=0 \nonumber
\end{eqnarray}
\begin{eqnarray}\label{11.10}
\mbox{and we have also} \quad (\Delta H)_{eq.} =0 \, .
\end{eqnarray}
We now note that from $(\ref{11.9})_{3,4}$ we deduce that the function $\frac{\partial
  \Delta H}{\partial \mu_{k}}$ has all the derivatives with respect to $\mu_i$, $\mu_{ij}$,
$\lambda_i$ which are symmetric tensors, so that its Taylor ' s expansion around
equilibrium is
\begin{eqnarray}\label{10.1}
&{}& \frac{\partial   \Delta H}{\partial  \mu_{k}}= \sum_{p,q,r}^{0 \cdots \infty}
\frac{1}{p!} \frac{1}{q!} \frac{1}{r!} H_{p+1,q,r}^{ki_1 \cdots i_ph_1k_1 \cdots
h_qk_qj_1 \cdots j_r} \mu_{i_1} \cdots \mu_{i_p}\mu_{h_1k_1} \cdots \mu_{h_qk_q}
\lambda_{j_1} \cdots \lambda_{j_r} \, ,
\end{eqnarray}
\begin{eqnarray*}
&{}& \mbox{where} \quad H_{p+1,q,r}^{ki_1 \cdots i_ph_1k_1 \cdots h_qk_qj_1 \cdots j_r} =
\left( \frac{\partial^{p+1+q+r}   \Delta H}{\partial \mu_{k} \partial \mu_{i_1} \cdots
\partial \mu_{i_p} \partial \mu_{h_1k_1} \cdots \partial \mu_{h_qk_q} \cdots  \partial  \lambda_{j_1} \cdots
\partial  \lambda_{j_r}} \right)_{eq.}
\end{eqnarray*}
is a symmetric tensor depending only on the scalars $\mu$ and $\lambda$, so that it has
the form
\begin{eqnarray*}
&{}& H_{p+1,q,r}^{ki_1 \cdots i_ph_1k_1 \cdots h_qk_qj_1 \cdots j_r} = \left\{
\begin{array}{ll}
  H_{p+1,q,r}(\mu ,\lambda) \delta^{(ki_1 \cdots i_ph_1k_1 \cdots h_qk_qj_1 \cdots j_r)} &\mbox{if $p+r+1$ is even}  \\
  0 & \mbox{if $p+r+1$ is odd}
\end{array} \right.
\end{eqnarray*}
where $\delta^{(a_1 \cdots a_{2n})}$ denotes $\delta^{(a_1 a_2} \cdots  \delta^{a_{2n-1}
a_{2n})}$. \\
By integrating (\ref{10.1}) we obtain
\begin{eqnarray}\label{10.2}
  \Delta H= \sum_{p,q}^{0 \cdots \infty} \sum_{r \in I_{p+1}} \frac{1}{(p+1)!} \frac{1}{q!}
\frac{1}{r!} H_{p+1,q,r} \delta^{(i_1 \cdots i_{p+1}h_1k_1 \cdots h_qk_qj_1 \cdots j_r)}
\mu_{i_1} \cdots \mu_{i_{p+1}}\mu_{h_1k_1} \cdots \mu_{h_qk_q} \lambda_{j_1} \cdots
\lambda_{j_r} + \\
+ \bar{H}(\mu , \mu_{ab} , \lambda , \lambda_c) \, , \nonumber
\end{eqnarray}
where $I_p$ denotes the set of all non negative integers $r$ such that $r+p$ is even. \\
But also $\frac{\partial   \Delta H}{\partial  \mu}$ has all the derivatives which are
symmetric tensors; in fact, the derivatives of $(\ref{11.9})_{3,4}$ with respect to $\mu$
are
\begin{eqnarray}\label{10.3}
&{}& \frac{\partial^3   \Delta H}{\partial  \mu \partial  \mu_{[k} \partial \mu_{i]j} } =
0 \quad , \quad \frac{\partial^3   \Delta H}{\partial  \mu \partial  \mu_{[k} \partial
\lambda_{i]} } = 0\, .
\end{eqnarray}
The derivatives of $(\ref{11.9})_{1,2}$ with respect to $\mu_{ab}$ are
\begin{eqnarray*}
&{}&  \frac{\partial^3   \Delta H}{\partial  \mu_{ab} \partial  \mu \partial  \mu_{ij}} =
\frac{\partial^3   \Delta H}{\partial  \mu_{ab} \partial \mu_i \partial  \mu_{j}} \quad ,
\quad \frac{\partial^3   \Delta  H}{\partial  \mu_{ab} \partial  \mu
\partial  \lambda_i} = \frac{\partial^3   \Delta H}{\partial  \mu_{ab} \partial \lambda \partial
\mu_{i}}\quad,
\end{eqnarray*}
whose skew-symmetric parts with respect to $b$ and $i$ are
\begin{eqnarray}\label{10.4}
&{}&  \frac{\partial^3   \Delta H}{\partial  \mu \partial  \mu_{a[b}  \partial
\mu_{i]j}} = 0
 \quad , \quad
\frac{\partial^3   \Delta H}{\partial  \mu \partial  \mu_{a[b}
\partial  \lambda_{i]}} = 0 \quad ,
\end{eqnarray}
thanks to eq. $(\ref{11.9})_{3,4}$. \\
By using (\ref{10.2}) we obtain that also $\frac{\partial \bar{H}}{\partial  \mu}$ has
all the derivatives which are symmetric tensors, so that its expansion around equilibrium
is
\begin{eqnarray}\label{11.1}
\frac{\partial \bar{H}}{\partial  \mu}= \sum_{q}^{0 \cdots \infty} \sum_{r \in I_{0}}
 \frac{1}{q!} \frac{1}{r!} \frac{\partial \bar{H}_{q,r}}{\partial \mu}(\mu , \lambda) \delta^{(h_1k_1
\cdots h_qk_qj_1 \cdots j_r)} \mu_{h_1k_1} \cdots \mu_{h_qk_q} \lambda_{j_1} \cdots
\lambda_{j_r}  \, ,
\end{eqnarray}
where the derivative of $\bar{H}_{q,r}$ has been introduced for later convenience and
without loss of generality. By integrating (\ref{11.1}) we obtain
\begin{eqnarray}\label{11.2}
\bar{H}= \sum_{q}^{0 \cdots \infty} \sum_{r \in I_{0}}
 \frac{1}{q!} \frac{1}{r!} \bar{H}_{q,r}(\mu , \lambda) \delta^{(h_1k_1
\cdots h_qk_qj_1 \cdots j_r)} \mu_{h_1k_1} \cdots \mu_{h_qk_q} \lambda_{j_1} \cdots
\lambda_{j_r} + \bar{\bar{H}}(\mu_{ab} , \lambda , \lambda_c) \, .
\end{eqnarray}
But the function $  \Delta H$ is present in (\ref{11.9}),  (\ref{8.2}) only through its
derivatives with respect to $\mu$ and $\mu_k$ so that the function $ \bar{\bar{H}}$
doesn't effect the results and, consequently, it is not restrictive to assume that it is
zero. By substituting (\ref{11.2}) in (\ref{10.2}) we see that the function $  \Delta H$
has derivatives which are symmetric tensors and we have also its expansion, that is
\begin{eqnarray}\label{10.5}
  \Delta H= \sum_{p,q}^{0 \cdots \infty} \sum_{r \in I_{p}} \frac{1}{p!} \frac{1}{q!} \frac{1}{r!}
H_{p,q,r}(\mu , \lambda) \delta^{(i_1 \cdots i_{p}h_1k_1 \cdots h_qk_qj_1 \cdots j_r)}
\mu_{i_1} \cdots \mu_{i_{p}}\mu_{h_1k_1} \cdots \mu_{h_qk_q} \lambda_{j_1} \cdots
\lambda_{j_r} \, ,
\end{eqnarray}
where we have defined $H_{0,q,r}=\bar{H}_{q,r}$; in this way, the term of (\ref{10.5})
with $p=0$ is (\ref{11.2}), while for the other terms we can change index according to
the law $p= 1+ P$ and obtain the remaining part of (\ref{10.2}). \\
Now  an interesting consequence of (\ref{11.9}) and
(\ref{11.10}) is that  \\
PROPERTY 1: " The expansion of $\Delta H$ up to order $n \geq 1$ with respect to
equilibrium is a polynomial of degree $n-1$ in the variable $\mu$." \\
Thanks to this property, it is not restrictive to assume for $\Delta H$ a polynomial
expansion of infty degree in the variable $\mu$; we simply expect that the equations will
stop by itself the terms with higher degree. \\
Consequently, it is not restrictive to assume that the functions $H_{p,q,r}$ in
(\ref{10.5}) can be written as
\begin{eqnarray}\label{10.50}
H_{p,q,r}= \sum_{s=0}^{\infty}  \frac{1}{s!} \mu^s H_{p,q,r,s}(\lambda)  \, .
\end{eqnarray}
So, even if $\mu$ is not zero at equilibrium, for what concerns $\Delta H$, we can do an
expansion also around $\mu=0$; obviously, the situation is different for the particular
solution $H_1$ reported in eq. (\ref{11.6}). We report the proof of Property 1, in
Appendix 2. \\
If we substitute (\ref{10.5}) and (\ref{10.50}) in $(\ref{11.9})_{3,4}$,  we obtain
identities. If we substitute (\ref{10.5}) and (\ref{10.50}) in $(\ref{11.9})_{1,2}$, we
obtain
\begin{eqnarray}\label{15.2}
H_{p,q+1,r,s+1} = H_{p+2,q,r,s} \quad , \quad H_{p,q,r+1,s+1} = \frac{\partial}{\partial
\lambda} H_{p+1,q,r,s} \, .
\end{eqnarray}
From $(\ref{15.2})_1$ we now obtain
\begin{eqnarray}\label{beta.2}
H_{p,q,r,s} = \left\{ \begin{array}{ll}
 H_{0,q+\frac{p}{2},r,s+\frac{p}{2}} & \mbox{if $p$ is even} \\
 H_{1,q+\frac{p-1}{2},r,s+\frac{p-1}{2}} & \mbox{if $p$ is odd} \, .
\end{array}
\right.
\end{eqnarray}
After that, we see that $(\ref{15.2})_1$ is satisfied as a consequence of (\ref{beta.2}).
\\
Let us focus now our attention to eq. $(\ref{15.2})_2$; for $p=0,1$ it becomes
\begin{eqnarray}\label{beta.3}
H_{0,q,r+1,s+1} = \frac{\partial}{\partial \lambda} H_{1,q,r,s} \quad , \quad
H_{1,q,r+1,s+1} = \frac{\partial}{\partial \lambda} H_{0,q+1,r,s+1} \, ,
\end{eqnarray}
where, for $(\ref{beta.3})_2$ we have used (\ref{beta.2}) with $p=2$. After that, eq.
$(\ref{15.2})_2$ with use of eq. (\ref{beta.2})
\begin{itemize}
  \item in the case with $p$ even, gives $(\ref{beta.3})_1$ with $(q+\frac{p}{2} ,
  s+\frac{p}{2})$ instead of $(q,s)$,
  \item in the case with $p$ odd, gives $(\ref{beta.3})_2$ with $(q+\frac{p-1}{2} ,
  s+\frac{p-1}{2})$ instead of $(q,s)$.
\end{itemize}
But we have now to impose $(\ref{11.9})_5$. To  this end, let us take its derivatives
with respect to  $\mu_{i_1}$, $\cdots$ , $\mu_{i_P}$, $\mu_{h_1k_1}$, $\cdots$ ,
$\mu_{h_Qk_Q}$, $\lambda_{j_1}$, $\cdots$ , $\lambda_{j_R}$ and let us calculate the
result at equilibrium; in this way we obtain
\begin{eqnarray*}
0 = P \delta^{i \overline{i_1}} \delta^{(\overline{i_2 \cdots i_{P}}kh_1k_1 \cdots
h_Qk_Qj_1 \cdots j_R)} H_{P,Q,R,s+1} +2 Q  \delta^{i \overline{h_1}}
\delta^{(\overline{k_1 h_2k_2 \cdots h_Qk_Q}ki_1 \cdots
i_{P}j_1 \cdots j_R)} H_{P,Q,R,s+1}+ \\
+ 2 \lambda   \delta^{(ki h_1k_1 \cdots h_Qk_Q i_{1} \cdots i_{P}j_1 \cdots j_R)}
H_{P,Q+1,R,s+1}
+ 2 R   \delta^{(ki h_1k_1 \cdots h_Qk_Q i_{1} \cdots i_{P}j_1 \cdots j_R)} H_{P+1,Q+1,R-1,s}+ \\
+R \delta^{i \overline{j_1}} \delta^{(\overline{j_2 \cdots j_{R}}kh_1k_1 \cdots
h_Qk_Qi_{1} \cdots i_{P})} H_{P,Q,R,s+1} +\delta^{ki} \delta^{(i_{1} \cdots i_{P} h_1k_1
\cdots h_Qk_Q j_1 \cdots j_R)} H_{P,Q,R,s+1} \, ,
\end{eqnarray*}
where overlined indexes denote symmetrization over those indexes, after that the other
one (round brackets around indexes) has been taken. (Note that, in the fourth term the
index $R-1$ appears; despite this fact, the equations holds also for $R=0$ but in this
case, this fourth term is not present as it is remembered also by the factor $R$).  \\
Now, the first, second, fifth and sixth term can be put together so that the above
expression becomes
\begin{eqnarray*}
0 = (P+2Q+R+1) H_{P,Q,R,s+1}  \delta^{i \overline{i_1}} \delta^{(\overline{i_2
\cdots i_{P}kh_1k_1 \cdots h_Qk_Qj_1 \cdots j_R})}  + \\
+ ( 2 \lambda H_{P,Q+1,R,s+1} + 2 R H_{P+1,Q+1,R-1,s})
 \delta^{(ki h_1k_1 \cdots h_Qk_Q i_{1} \cdots i_{P}j_1 \cdots j_R)} \, ,
\end{eqnarray*}
that is
\begin{eqnarray}\label{17.1}
0 = (P+2Q+R+1) H_{P,Q,R,s+1} + 2 \lambda H_{P,Q+1,R,s+1} + 2 R H_{P+1,Q+1,R-1,s} \, .
\end{eqnarray}
This relation, for $P=0,1$ reads
\begin{eqnarray}\label{beta.4}
&{}& 0 = (2Q+R+1) H_{0,Q,R,s+1} + 2 \lambda H_{0,Q+1,R,s+1} + 2 R
H_{1,Q+1,R-1,s}  \, , \\
&{}& 0 = (2Q+R+2)H_{1,Q,R,s+1} + 2 \lambda H_{1,Q+1,R,s+1} + 2 R H_{0,Q+2,R-1,s+1}  \, ,
\nonumber
\end{eqnarray}
with the agreement that the last terms are not present in the case $R=0$.  (For eq.
$(\ref{beta.4})_2$ we have used $(\ref{beta.2})$ with $p=2$).  \\
For the other values of $P$, eq. $(\ref{17.1})$ with use of eq. (\ref{beta.2})
\begin{itemize}
  \item in the case with $p$ even, gives $(\ref{beta.4})_1$ with $(q+\frac{p}{2} ,
  s+\frac{p}{2})$ instead of $(q,s)$,
  \item in the case with $p$ odd, gives $(\ref{beta.4})_2$ with $(q+\frac{p-1}{2} ,
  s+\frac{p-1}{2})$ instead of $(q,s)$.
\end{itemize}
Summarizing the results, we have that (\ref{beta.2}) gives $H_{P,Q,R,s}$ in terms of
$H_{0,Q,R,s}$ and $H_{1,Q,R,s}$, while eqs. (\ref{beta.3}) and (\ref{beta.4}) give
restrictions on $H_{0,Q,R,s}$ and $H_{1,Q,R,s}$. Let us now see how to solve these
restrictions. First of all, eq. (\ref{11.10}) expressed in terms of (\ref{10.5}) and
(\ref{10.50}) becomes
\begin{eqnarray}\label{cris1}
&{}& H_{0,0,0,s} =0  \quad \forall \, s \, .
\end{eqnarray}
We have also already imposed that $\Delta H$ becomes zero when calculated in $\mu=0$ and
$\mu_k=0$. This can be expressed in terms of (\ref{10.5}) and (\ref{10.50}) as
\begin{eqnarray}\label{cris2}
&{}& H_{0,q,r,0} =0  \quad \forall \, q, r \, .
\end{eqnarray}
From $(\ref{beta.4})_1$ with $R=0$, by using (\ref{cris1}) and (\ref{cris2}), we find
\begin{eqnarray}\label{cris3}
&{}& H_{0,q,0,s} =0  \quad \forall \, q, s \, .
\end{eqnarray}
Let us now apply $(\ref{beta.3})_2$ with $(r-1,s-1)$ instead of $(r,s)$ and,
subsequently, $(\ref{beta.3})_1$, so obtaining
\begin{eqnarray}\label{cris4}
&{}& H_{1,q,r,s} = \frac{\partial^{2}}{\partial \lambda^{2}} H_{1,q+1,r-2,s-1} \quad
\mbox{from which} \quad H_{1,q,r,s} = \frac{\partial^{2k}}{\partial \lambda^{2k}}
H_{1,q+k,r-2k,s-k}
\end{eqnarray}
where the values of $k$ are restricted by $2k \leq r$ and $k \leq s$. So, if $s \leq
\frac{r-1}{2}$ ( we recall that $r$ is odd because the sum of the first and third index
must be an even number), we can take $k=s$ so obtaining $H_{1,q,r,s}$ in terms of
$H_{1,q,r,0}$; if $s \geq \frac{r+1}{2}$ we take $k = \frac{r-1}{2}$ so obtaining
$H_{1,q,r,s}$ in terms of $H_{1,q+\frac{r-1}{2},1,s-\frac{r-1}{2}}$ which is zero for
$(\ref{beta.3})_2$ and (\ref{cris3}). So we have obtained that
\begin{eqnarray}\label{cris5}
&{}& H_{1,q,r,s} = \left\{  \begin{array}{cc}
  \frac{\partial^{2s}}{\partial \lambda^{2s}} H_{1,q+s,r-2s,0}  & \mbox{if} \quad s \leq \frac{r-1}{2} \\
  & \\
  0 & \mbox{if} \quad s \geq \frac{r+1}{2}
\end{array}
\right.
\end{eqnarray}
By applying this result, $(\ref{beta.3})_1$ becomes
\begin{eqnarray}\label{cris6}
&{}& H_{0,q,r,s+1} = \left\{  \begin{array}{cc} 0 & \mbox{if} \quad r=0 \\
  & \\
  \frac{\partial^{2s+1}}{\partial \lambda^{2s+1}} H_{1,q+s,r-2s-1,0}  & \mbox{if} \quad s \leq \frac{r-2}{2}
  \quad , \quad  r \geq 2\\
  & \\
  0 & \mbox{if} \quad s \geq \frac{r}{2} \quad , \quad r \geq 2
\end{array}
\right.
\end{eqnarray}
where, for the value with $r=0$ we have used (\ref{cris3}). So every thing is determined
in terms of $H_{1,q,r,0}$ which remains arbitrary up to now. After that
$(\ref{beta.3})_{1,2}$ become  consequences of (\ref{cris5}) and (\ref{cris6}). \\
Regarding $(\ref{beta.4})_1$, we have already imposed it for $R=0$; if $R \geq 2$, $s
\geq \frac{R}{2}$  it is an identity, as a consequence of (\ref{cris5}) and
(\ref{cris6}). But if $R \geq 2$, $s \leq \frac{R-2}{2}$  by using (\ref{cris5}) and
(\ref{cris6}) it becomes
\begin{eqnarray}\label{cris7}
 0 = (2Q+R+1) \frac{\partial}{\partial \lambda} H_{1,Q,R-1,0} + 2 \lambda
\frac{\partial}{\partial \lambda} H_{1,Q+1,R-1,0} + 2 R H_{1,Q+1,R-1,0}  \, , \quad
\mbox{for} \quad  R \geq 2 \, ,
\end{eqnarray}
for $s=0$, while for $s \geq 1$ it becomes
\begin{eqnarray*}
0 = (2Q+R+1) \frac{\partial^{2s+1}}{\partial \lambda^{2s+1}} H_{1,Q+s,R-2s-1,0} + 2
\lambda \frac{\partial^{2s+1}}{\partial \lambda^{2s+1}} H_{1,Q+1+s,R-1-2s,0} + 2 R
\frac{\partial^{2s}}{\partial \lambda^{2s}} H_{1,Q+1+s,R-1-2s,0}
\end{eqnarray*}
and this is $\frac{\partial^{2s}}{\partial \lambda^{2s}}$ of the relation obtained from
(\ref{cris7}) with $(Q+s, R-2s)$ instead of $(Q, R)$. \\
Regarding $(\ref{beta.4})_2$, for $R \geq 3$, $s \geq \frac{R-1}{2}$  it is an identity,
as a consequence of (\ref{cris5}) and (\ref{cris6}). Similarly, if $s \leq
\frac{R-3}{2}$, $R \geq 3$ by using (\ref{cris5}) and (\ref{cris6}) it becomes
\begin{eqnarray*}
0 = (2q+r+2) \frac{\partial^{2s+2}}{\partial \lambda^{2s+2}} H_{1,q+s+1,r-2s-2,0} + 2
\lambda \frac{\partial^{2s+2}}{\partial \lambda^{2s+2}} H_{1,q+2+s,r-2-2s,0} + 2 r
\frac{\partial^{2s+1}}{\partial \lambda^{2s+1}} H_{1,q+2+s,r-2-2s,0}
\end{eqnarray*}
and this is $\frac{\partial^{2s+1}}{\partial \lambda^{2s+1}}$ of the relation obtained
from (\ref{cris7}) with $Q= q+s+1$, $R= r-2s-1$. \\
Finally, in the case $R=1$ by using (\ref{cris5}), (\ref{cris6}) and (\ref{cris3}) it
becomes an identity.  We note that (\ref{cris7}) can be written as
\begin{eqnarray}\label{cris8}
 0 = (2Q+R+1) \frac{\partial}{\partial \lambda} H_{1,Q,R-1,0} + 2 \lambda^{1-R}
\frac{\partial}{\partial \lambda} \left( \lambda^{R} H_{1,Q+1,R-1,0} \right)   \quad
\mbox{for} \quad  R \geq 2 \, ,
\end{eqnarray}
so that it allows to determine, with an integration, the function $H_{1,Q+1,R-1,0}$ in
terms of $H_{1,Q,R-1,0}$. So $H_{1,0,R-1,0}$ remains arbitrary and also the family of
constants arising from the integrations.

\section{A second order theory}
For a second order theory, we need the expressions of $h'$ and $h'^k$ up to third order.
By using the results of the previous sections, we can see that they are determined in
terms of $\psi_1$, $H_{1,0,1,0}$, $H_{1,1,1,0}$, $H_{1,2,1,0}$, $H_{1,0,3,0}$  and they
are
\begin{eqnarray}\label{cris9}
&{}& h'= h_{0,0,0} + h_{0,1,0} \mu_{ll} + \\
&{}& \quad \quad +\frac{1}{2} h_{2,0,0} \mu_{i} \mu^{i} + h_{1,0,1} \mu_{i} \lambda^{i} +
     \frac{1}{2}  h_{0,2,0} \delta^{(ab}  \delta^{cd)}  \mu_{ab} \mu_{cd} +
     \quad \quad  \frac{1}{2}  h_{0,0,2} \lambda^{i} \lambda_{i} + \nonumber \\
&{}&  \quad \quad    +  \frac{1}{2}  h_{2,1,0}  \delta^{(ab}  \delta^{cd)} \mu_{a}
\mu_{b} \mu_{cd} +
     h_{1,1,1}  \delta^{(ab}  \delta^{cd)} \mu_{a} \lambda_{b} \mu_{cd} +
     \frac{1}{6} h_{0,3,0}   \delta^{(a_1a_2}  \delta^{a_3a_4}   \delta^{a_5a_6)} \mu_{a_1a_2} \mu_{a_3a_4}
     \mu_{a_5a_6}+   \nonumber\\
&{}&  \quad \quad    +  \frac{1}{2}  h_{0,1,2}  \delta^{(ab}  \delta^{cd)} \lambda_{a}
\lambda_{b} \mu_{cd}  \nonumber \, ,
\end{eqnarray}
\begin{eqnarray}\label{cris10}
&{}&  h'^k= \phi_{1,0,0} \mu^{k}+ \phi_{0,0,1} \lambda^{k} + \\
&{}&     \quad \quad   + \phi_{1,1,0}  \delta^{(ka}  \delta^{bc)} \mu_{a} \mu_{bc} +
     \phi_{0,1,1}  \delta^{(ka}  \delta^{bc)} \lambda_{a} \mu_{bc} +
     \frac{1}{6} \phi_{3,0,0}  \delta^{(ka}  \delta^{bc)} \mu_{a} \mu_{b} \mu_{c} +
     \frac{1}{2}  \phi_{2,0,1}  \delta^{(ka}  \delta^{bc)} \mu_{a} \mu_{b} \lambda_{c} + \nonumber \\
 &{}&    \quad \quad   + \frac{1}{2} \phi_{1,2,0}   \delta^{(ka}  \delta^{bc}   \delta^{de)} \mu_{a} \mu_{bc} \mu_{de}+
     \frac{1}{2} \phi_{1,0,2}  \delta^{(ka}  \delta^{bc)} \mu_{a} \lambda_{b} \lambda_{c} +
      \frac{1}{2} \phi_{0,2,1}  \delta^{(ka}  \delta^{bc}   \delta^{de)} \mu_{ab} \mu_{cd} \lambda_{e}
      + \nonumber \\
 &{}&  \quad \quad  + \frac{1}{6} \phi_{0,0,3}  \delta^{(ka}  \delta^{bc)} \lambda_{a} \lambda_{b}
 \lambda_{c} \nonumber \, ,
\end{eqnarray}
with
\begin{eqnarray}\label{cris11}
&{}& h_{0,0,0} = \frac{\partial^2}{\partial \mu^2} \psi_1 \quad , \quad
h_{0,1,0}=\frac{\partial^2}{\partial \mu^2}   \left( \frac{-1}{2 \lambda} \psi_1 \right)
\quad , \quad h_{2,0,0}= \frac{\partial^3}{\partial \mu^3}   \left( \frac{-1}{2 \lambda}
\psi_1 \right) \, , \\
&{}& h_{1,0,1}=\frac{\partial^3}{\partial \lambda \partial \mu^2}   \left( \frac{-1}{2
\lambda} \psi_1 \right)  \quad , \quad h_{0,2,0} = 3\frac{\partial^2}{\partial \mu^2}
\left[ \left( \frac{-1}{2 \lambda} \right)^{2} \psi_{1} \right]  \quad ,  \nonumber \\
&{}& h_{0,0,2}= \frac{\partial^3}{\partial \lambda^2 \partial \mu}   \left( \frac{-1}{2
\lambda} \psi_1 \right) + \frac{\partial}{\partial \lambda } H_{1,0,1,0}  \quad , \quad
h_{2,1,0}= 3\frac{\partial^3}{\partial \mu^3}   \left[ \left( \frac{-1}{2 \lambda}
\right)^{2} \psi_{1} \right]  \, , \nonumber \\
&{}&h_{1,1,1}= 3\frac{\partial^3}{\partial \lambda  \partial \mu^2} \left[ \left(
\frac{-1}{2 \lambda} \right)^{2} \psi_{1} \right]  \quad , \quad h_{0,3,0}=15
\frac{\partial^2}{\partial \mu^2}   \left[ \left(
\frac{-1}{2 \lambda} \right)^{3} \psi_{1} \right] \, ,  \nonumber \\
&{}& h_{0,1,2}= 3\frac{\partial^3}{\partial \lambda^2  \partial \mu}   \left[ \left(
\frac{-1}{2 \lambda} \right)^{2} \psi_{1} \right] + \frac{\partial}{\partial \lambda }
H_{1,1,1,0} \, ,  \nonumber
\end{eqnarray}
\begin{eqnarray}\label{cris12}
&{}&  \phi_{1,0,0} =\frac{\partial^2}{\partial \mu^2}   \left( \frac{-1}{2 \lambda}
\psi_1 \right) \quad , \quad  \phi_{0,0,1}= \frac{\partial^2}{\partial \lambda \partial
\mu}   \left( \frac{-1}{2 \lambda} \psi_1 \right) + H_{1,0,1,0}  \, , \\
&{}&  \phi_{1,1,0} = 3\frac{\partial^2}{\partial \mu^2} \left[ \left( \frac{-1}{2
\lambda} \right)^{2} \psi_{1} \right]  \quad , \quad \phi_{0,1,1}=
3\frac{\partial^2}{\partial \lambda  \partial \mu} \left[ \left( \frac{-1}{2 \lambda}
\right)^{2} \psi_{1} \right] + H_{1,1,1,0}  \quad ,  \nonumber \\
&{}&  \phi_{3,0,0}= 3\frac{\partial^3}{\partial \mu^3}   \left[ \left( \frac{-1}{2
\lambda} \right)^{2} \psi_{1} \right]  \quad , \quad \phi_{2,0,1}=
3\frac{\partial^3}{\partial \lambda  \partial \mu^2} \left[ \left( \frac{-1}{2 \lambda}
\right)^{2} \psi_{1} \right] \quad ,  \nonumber \\
&{}&  \phi_{1,2,0}=15 \frac{\partial^2}{\partial \mu^2}   \left[ \left( \frac{-1}{2
\lambda} \right)^{3} \psi_{1} \right]  \quad , \quad
\phi_{1,0,2}=3\frac{\partial^3}{\partial \lambda^2  \partial \mu}   \left[ \left(
\frac{-1}{2 \lambda} \right)^{2} \psi_{1} \right] + \frac{\partial}{\partial \lambda }
H_{1,1,1,0}  \quad ,   \nonumber \\
&{}&  \phi_{0,2,1}= 15 \frac{\partial^2}{\partial \lambda  \partial \mu}   \left[ \left(
\frac{-1}{2 \lambda} \right)^{3} \psi_{1} \right] + H_{1,2,1,0}  \quad ,  \nonumber \\
&{}& \phi_{0,0,3}= 3 \frac{\partial^3}{\partial \lambda^3}   \left[ \left( \frac{-1}{2
\lambda} \right)^{2} \psi_{1} \right] + H_{1,0,3,0} + \mu \frac{\partial^2}{\partial
\lambda^2 }  H_{1,1,1,0}  \quad .  \nonumber
\end{eqnarray}
Here $\psi_1(\mu , \lambda)$, $H_{1,0,1,0}(\lambda)$,  $H_{1,0,3,0}(\lambda)$ are
arbitrary functions while  $H_{1,1,1,0}$ and $H_{1,2,1,0}$ are linked to them trough
(\ref{cris8}) which, for their case, becomes
\begin{eqnarray}\label{cris12a}
&{}&  \frac{\partial}{\partial \lambda} \left( \lambda^{2} H_{1,1,1,0} \right) =
-\frac{3}{2}  \, \lambda  \,  \frac{\partial}{\partial \lambda} H_{1,0,1,0}   \quad ,
\quad \frac{\partial}{\partial \lambda} \left( \lambda^{2} H_{1,2,1,0} \right) =
-\frac{5}{2} \, \lambda \,  \frac{\partial}{\partial \lambda} H_{1,1,1,0}  \, .
\end{eqnarray}
Now we want express our closure by taking the moments as independent variables. To this
end we will indicate with $()_{eq.}$ the quantity $()$ calculated at equilibrium when
this is defined as the state with $\mu_i=0$, $\mu_{ij}=0$, $\lambda_i=0$; we will
indicate with $()_{Eq.}$ the quantity $()$ calculated at equilibrium when this is defined
as the state with , $\hat{F}^{ij}=0$, $\hat{G}^{i}=0$. Both states are coincident at the
order zero.

\subsection{The closure at equilibrium}
Let us substitute (\ref{cris9}) and (\ref{cris10}) in $(\ref{2bis.3})_1$ and calculate
the right hand sides at equilibrium and the left hand side at Equilibrium and for
$v^i=0$. So we obtain
\begin{eqnarray}\label{cris13}
\hat{F} = \rho = \frac{\partial h_{0,0,0}}{\partial \mu}  \quad , \quad \hat{F}^{ij} = p
\delta^{ij} =  h_{0,1,0}  \delta^{ij} \quad , \quad
 \hat{G} = 2 \rho \varepsilon = \frac{\partial h_{0,0,0}}{\partial \lambda}  \quad ,
\quad \hat{G}^{ill} = 0 \, ,
\end{eqnarray}
where $p$ is the pressure and $\varepsilon$ the internal energy density. The same
procedure applied to $(\ref{2bis.3})_2$ and $(\ref{2bis.2})_1$ gives
\begin{eqnarray}\label{cris14}
&{}&  \hat{F}^{kij} = 0 \, , \, \hat{G}^{kill} =  \phi_{0,0,1}\delta^{ik} \, , \\
&{}&  h_{Eq.} = - h'_{eq.} + \mu \rho + \lambda 2 \rho \varepsilon = -  h_{0,0,0} + \mu
\frac{\partial h_{0,0,0}}{\partial \mu} + \lambda \frac{\partial h_{0,0,0}}{\partial
\lambda} \, .  \nonumber
\end{eqnarray}
From the second of these expressions and from the definition $h_{Eq.}= \rho s$, we obtain
\\
$d \, s - 2 \lambda [ d \, \varepsilon + p d \, (1/ \rho)]= 0$, where we have used
$h_{0,1,0}= - \frac{1}{2\lambda} h_{0,0,0}$ which comes out from $(\ref{cris11})_{1,2}$.
The result, thanks to the Gibbs Relation, allows to identify as in \cite{1} that
\begin{eqnarray}\label{cris15}
&{}&  \lambda_{Eq.} = \frac{1}{2T} \, ,
\end{eqnarray}
with $T$ the absolute temperature. We will see in the sequel that, if we take as
variables describing equilibrium the absolute temperature $T$ and the chemical potential
$\mu_{Eq.}$, then all our closure will be determined in terms of the arbitrary functions
$\psi_1(\mu , \lambda)$, $H_{1,0,1,0}(\lambda)$,  $H_{1,0,3,0}(\lambda)$ and of the two
constants arising from integration of (\ref{cris12a}). Also $p$, $\rho$ and $\varepsilon$
will be functions depending on them; in particular, for a third order theory we will have
\begin{eqnarray}\label{p.2q}
&{}&  \rho = \frac{\partial^3}{\partial \mu^3} \psi_1 \quad , \quad p= -T
\frac{\partial^2}{\partial \mu^2} \psi_1 \quad , \quad 2 \rho \varepsilon = -2T^2
\frac{\partial^3}{\partial T \partial  \mu^2} \psi_1 \, .
\end{eqnarray}
Instead of this, if we take as variables describing equilibrium $\rho$ and $T$ we will
see that more difficult calculations are needed. To reach this end, let us now in the
remaining part of this
subsection to prepare some results which will be used later. \\
Let us use $(\ref{cris13})_{1}$ to find $\mu_{Eq.}$ as a function of $(\rho,T)$. By
taking its derivatives  with respect to $\rho$ and $T$, we obtain
\begin{eqnarray}\label{cris16}
\frac{\partial \mu}{\partial \rho} = \left( \frac{\partial^4 \psi_1}{\partial \mu^4}
\right)^{-1} \quad , \quad \frac{\partial \mu}{\partial T} = \frac{1}{2} \left(
\frac{\partial^4 \psi_1}{\partial \mu^4} \right)^{-1} \frac{\partial^4 \psi_1}{\partial
\lambda \partial \mu^3} \frac{1}{T^2} \, .
\end{eqnarray}
Similarly, by taking the derivatives  with respect to $\rho$ and $T$ of
\begin{eqnarray}\label{cris17}
p= - \frac{1}{2\lambda} \frac{\partial^2 \psi_1}{\partial \mu^2} = -T h_{0,0,0} \, ,
\end{eqnarray}
we find
\begin{eqnarray}\label{cris18}
 \left(\frac{\partial p}{\partial \rho}\right)_{T} = - \frac{1}{2\lambda}
\frac{\partial^3 \psi_1}{\partial \mu^3} \frac{\partial \mu}{\partial \rho} = -T
\frac{\partial^3 \psi_1}{\partial \mu^3} \left( \frac{\partial^4 \psi_1}{\partial \mu^4}
\right)^{-1} = -\rho T \left( \frac{\partial^4 \psi_1}{\partial \mu^4} \right)^{-1} \, ,
\quad \quad \quad \quad \quad \quad \quad \quad \quad \quad
\end{eqnarray}
\begin{eqnarray*}
\left(\frac{\partial p}{\partial T}\right)_{\rho} = - \frac{1}{2T} \frac{\partial^3
\psi_1}{\partial \mu^3} \left( \frac{\partial^4 \psi_1}{\partial \mu^4} \right)^{-1}
\frac{\partial^4 \psi_1}{\partial \lambda \partial \mu^3} - \frac{1}{2} \frac{\partial
}{\partial \lambda}\left( \frac{-1}{2\lambda} \frac{\partial^2 \psi_1}{\partial
\mu^2}\right) \frac{1}{T^2} = \frac{p+\rho
\varepsilon}{T}-\frac{\rho}{2T}\frac{\partial^4 \psi_1}{\partial \lambda \partial \mu^3}
\left( \frac{\partial^4 \psi_1}{\partial \mu^4} \right)^{-1} \, .
\end{eqnarray*}
By taking the derivatives  with respect to $\rho$ and $T$ of
\begin{eqnarray}\label{cris19}
\varepsilon= \frac{1}{2} \left( \frac{\partial^3 \psi_1}{\partial \mu^3} \right)^{-1}
\frac{\partial^3 \psi_1}{\partial \lambda \partial \mu^2} = \frac{1}{2 \rho}
\frac{\partial h_{0,0,0}}{\partial \lambda} \, ,
\end{eqnarray}
we find
\begin{eqnarray}\label{cris20}
&{}& \left(\frac{\partial \varepsilon}{\partial \rho}\right)_{T} =  \frac{1}{2} \left[ -
  \left( \frac{\partial^3 \psi_1}{\partial \mu^3} \right)^{-2} \frac{\partial^4 \psi_1}{\partial \mu^4}
  \frac{\partial^3 \psi_1}{\partial \lambda \partial \mu^2} +
  \left( \frac{\partial^3 \psi_1}{\partial \mu^3} \right)^{-1}
  \frac{\partial^4 \psi_1}{\partial \lambda \partial \mu^3} \right]
  \left( \frac{\partial^4 \psi_1}{\partial \mu^4} \right)^{-1} = \\
&{}& \quad \quad \quad \quad \quad \quad=  - \frac{\varepsilon}{\rho} + \frac{1}{2 \rho}
\frac{\partial^4 \psi_1}{\partial \lambda \partial \mu^3} \left( \frac{\partial^4
\psi_1}{\partial \mu^4} \right)^{-1} \, . \nonumber
\end{eqnarray}
\begin{eqnarray*}
&{}& \left(\frac{\partial \varepsilon}{\partial T}\right)_{\rho} =  \frac{1}{2} \left[ -
  \left( \frac{\partial^3 \psi_1}{\partial \mu^3} \right)^{-2}
  \frac{\partial^4 \psi_1}{\partial \lambda \partial\mu^3}
  \frac{\partial^3 \psi_1}{\partial \lambda \partial \mu^2} +
  \left( \frac{\partial^3 \psi_1}{\partial \mu^3} \right)^{-1}
  \frac{\partial^4 \psi_1}{\partial \lambda^2 \partial \mu^2} \right] \frac{-1}{2T^2} +
  \\
&{}&\quad \quad \quad \quad   +  \frac{1}{2} \left[ -
  \left( \frac{\partial^3 \psi_1}{\partial \mu^3} \right)^{-2}
  \frac{\partial^4 \psi_1}{ \partial\mu^4}
  \frac{\partial^3 \psi_1}{\partial \lambda \partial \mu^2} +
  \left( \frac{\partial^3 \psi_1}{\partial \mu^3} \right)^{-1}
  \frac{\partial^4 \psi_1}{\partial \lambda \partial \mu^3} \right] \frac{1}{2T^2}
   \left( \frac{\partial^4 \psi_1}{\partial \mu^4} \right)^{-1}
   \frac{\partial^4 \psi_1}{\partial \lambda \partial \mu^3}= \\
&{}& \quad \quad \quad \quad  = \frac{-1}{4 \rho T^2}\frac{\partial^4 \psi_1}{\partial
\lambda ^2\partial \mu^2} +
 \frac{1}{4 \rho T^2} \left( \frac{\partial^4 \psi_1}{\partial \lambda \partial \mu^3} \right)^{2}
\left( \frac{\partial^4 \psi_1}{\partial \mu^4} \right)^{-1} \, .
\end{eqnarray*}
Well, from $(\ref{cris18})$ and $(\ref{cris20})_{2}$ we deduce
\begin{eqnarray}\label{cris21}
&{}& \left( \frac{\partial^4 \psi_1}{\partial \mu^4} \right)_{Eq.} = - \rho T
\left(\frac{\partial p}{\partial \rho}\right)_{T}^{-1} \, , \\
&{}& \left( \frac{\partial^4 \psi_1}{\partial \lambda \partial \mu^3} \right)_{Eq.} = 2 T
\left[ T \left(\frac{\partial p}{\partial T}\right)_{\rho} -p - \rho \varepsilon \right]
\left(\frac{\partial p}{\partial \rho}\right)_{T}^{-1} \, ,  \nonumber \\
&{}& \left( \frac{\partial^4 \psi_1}{\partial \lambda^2 \partial \mu^2} \right)_{Eq.} = -
4 \rho T^2 \left(\frac{\partial \varepsilon}{\partial T}\right)_{\rho} - \frac{4T}{\rho}
\left[ T \left(\frac{\partial p}{\partial T}\right)_{\rho} -p - \rho \varepsilon
\right]^2 \left(\frac{\partial p}{\partial \rho}\right)_{T}^{-1} \, .  \nonumber
\end{eqnarray}
By substituting in $(\ref{cris20})_{1}$ we obtain
\begin{eqnarray}\label{cris22}
&{}&  \left(\frac{\partial \varepsilon}{\partial \rho}\right)_{T} = \frac{1}{\rho^2}
\left[ p - T \left(\frac{\partial p}{\partial T}\right)_{\rho} \right]
\end{eqnarray}
which is well known because from the Gibbs Relation it follows
\begin{eqnarray}\label{cris23}
&{}&  \left(\frac{\partial s}{\partial T}\right)_{\rho} = \frac{1}{T}
\left(\frac{\partial \varepsilon}{\partial T}\right)_{\rho} \quad , \quad
\left(\frac{\partial s}{\partial \rho}\right)_{T} =  \frac{1}{T} \left(\frac{\partial
\varepsilon}{\partial \rho}\right)_{T} \, - \, \frac{p}{\rho^2 T} \, ,
\end{eqnarray}
whose integrability condition is just (\ref{cris22}). Eqs. (\ref{cris16}), thanks to
(\ref{cris21}), now become
\begin{eqnarray}\label{cris24}
\frac{\partial \mu}{\partial \rho} = -  \frac{1}{\rho T} \frac{\partial p}{\partial \rho}
\quad , \quad \frac{\partial \mu}{\partial T} = -  \frac{1}{\rho T^2} \left[ T
\left(\frac{\partial p}{\partial T}\right)_{\rho} -p - \rho \varepsilon \right] \, .
\end{eqnarray}
Regarding the closure for the fluxes at first order, we have found
$(\ref{cris14})_{1,2}$, where the coefficient $\phi_{0,0,1}$ appears. This is given by
$(\ref{cris12})_{2}$; to find a relation between it and $h_{0,0,0}$ given by
$(\ref{cris11})_{1}$, let us take the derivative
\begin{eqnarray*}
\frac{\partial \phi_{0,0,1}}{\partial \rho} = \frac{\partial \phi_{0,0,1}}{\partial
\lambda} \, \frac{\partial \lambda}{\partial \rho} + \frac{\partial
\phi_{0,0,1}}{\partial \mu} \, \frac{\partial \mu}{\partial \rho} = \frac{\partial
\phi_{0,0,1}}{\partial \mu} \, \frac{\partial \mu}{\partial \rho} = \left[ \frac{\partial
}{\partial \lambda} \left( -  \frac{1}{2 \lambda} \frac{\partial^2 \psi_1 }{ \partial
\mu^2} \right) \right] \frac{\partial \mu}{\partial \rho} = \left[ \frac{\partial
}{\partial \lambda} \left( -  \frac{1}{2 \lambda}  h_{0,0,0} \right) \right]
\frac{\partial \mu}{\partial \rho} \, .
\end{eqnarray*}
By using also $(\ref{cris16})_{1}$,  $(\ref{cris19})$, $(\ref{cris17})$,
$(\ref{cris13})_1$ and $(\ref{cris18})$, we obtain
\begin{eqnarray}\label{cris25}
\frac{\partial \phi_{0,0,1}}{\partial \rho} = 2 \left( \varepsilon + \frac{p}{\rho}
\right)  \left(\frac{\partial p}{\partial \rho}\right)_{T}  \, .
\end{eqnarray}
By comparing eqs. (\ref{cris14}) of the present article with eqs. (47) of \cite{1}
calculated at equilibrium, we find the same result with
\begin{eqnarray}\label{cris26}
\beta_1 = \left( \phi_{0,0,1} \right)_{Eq.} \, .
\end{eqnarray}
Moreover, in \cite{1} there was only the condition $(44)_1$ on $\beta_1$ (because
$(44)_4$ can be considered as the definition of $h_4$) and this condition is exactly the
present condition (\ref{cris25}). Consequently, we also can define $h_4$ from
\begin{eqnarray}\label{cris27}
\frac{\partial \phi_{0,0,1}}{\partial T}= 2 \left( \varepsilon + \frac{p}{\rho} \right)
\left(\frac{\partial p}{\partial T}\right)_{\rho} - \frac{h_4}{T^2}\, .
\end{eqnarray}
Making the point of situation, at equilibrium we have 3 state functions describing the
particular material under consideration: $p(\rho , T)$,  $\varepsilon(\rho , T)$,
$\beta_1(\rho , T)$. From the mathematical viewpoint they are arbitrary except that the
first 2 of them are linked by (\ref{cris22}) and the third one is determined by
(\ref{cris25}), except for an arbitrary function of the single variable $T$. In the
sequel, we will always use the identification (\ref{cris26}). \\
Other expressions which will be useful in the sequel are the following ones
\begin{eqnarray}\label{cris28}
\left( \frac{\partial h_{0,1,0}}{\partial \mu} \right)_{Eq.} = - \rho T \, ; \\
\left( \frac{\partial h_{0,1,0}}{\partial \lambda} \right)_{Eq.} = \frac{\partial^3
}{\partial \lambda \partial \mu^2} \left( -  \frac{1}{2 \lambda}  \psi_1 \right)=
\frac{\partial \phi_{0,0,1}}{\partial \mu} = \frac{\partial }{\partial \lambda} \left( -
\frac{1}{2 \lambda}  h_{0,0,0} \right) = \nonumber \\
=\frac{1}{2 \lambda^2} h_{0,0,0} -  \frac{1}{2 \lambda} \frac{\partial
h_{0,0,0}}{\partial \lambda} \stackrel{(1)}{=} -2Tp -2T \rho \varepsilon = -2T \rho
\left( \varepsilon + \frac{p}{\rho} \right) \, ; \nonumber \\
\left( h_{0,2,0} \right)_{Eq.} = 3 \frac{\partial^2 }{\partial \mu^2} \left[ \left( -
\frac{1}{2 \lambda} \right)^2 \psi_1 \right] \stackrel{(2)}{=} - 3 p T \, . \nonumber
\end{eqnarray}
These relations are consequences of (\ref{cris11}), (\ref{cris13}), $(\ref{cris12})_2$
and (\ref{cris26}); moreover, in the passage denoted with $\stackrel{(1)}{=}$ we have
used eq. (\ref{cris17}) and (\ref{cris19}), while in the passage denoted with
$\stackrel{(2)}{=}$ we have used eq. (\ref{cris17}). \\
We can now proceed and note that from (\ref{cris21}) it follows that
\begin{eqnarray}\label{cris29}
\left| \begin{array}{cc}
  \left( \frac{\partial^2 h_{0,0,0}}{\partial \mu^2} \right)_{Eq.}  &
  \left( \frac{\partial^2 h_{0,0,0}}{\partial \lambda \partial \mu} \right)_{Eq.}  \\
  & \\
  \left( \frac{\partial^2 h_{0,0,0}}{\partial \lambda \partial \mu} \right)_{Eq.} &
  \left( \frac{\partial^2 h_{0,0,0}}{\partial \lambda^2} \right)_{Eq.}
\end{array} \right| =
4 \rho^2 T^3 \frac{\partial \varepsilon}{\partial T} \left( \frac{\partial p }{\partial
\rho} \right)^{-1}
\end{eqnarray}
and that the determinant $D$ of the matrix
\begin{eqnarray}\label{cris30}
\left| \begin{array}{ccc}
  \left( \frac{\partial^2 h_{0,0,0}}{\partial \mu^2} \right)_{Eq.}  &
  \left( \frac{\partial^2 h_{0,0,0}}{\partial \lambda \partial \mu} \right)_{Eq.}  & - \rho T \\
  & & \\
  \left( \frac{\partial^2 h_{0,0,0}}{\partial \lambda \partial \mu} \right)_{Eq.} &
  \left( \frac{\partial^2 h_{0,0,0}}{\partial \lambda^2} \right)_{Eq.} &  -2 T (p+ \rho
  \varepsilon)  \\
  &  &  \\
  - \rho T  & -2 T (p+ \rho
  \varepsilon)   & - \frac{5}{3}  p T
\end{array} \right|
\end{eqnarray}
is equal to
\begin{eqnarray}\label{cris31}
&{}& D= 8 \rho^2 T^3 \frac{\partial \varepsilon}{\partial T} h_2 \left( \frac{\partial p
}{\partial \rho} \right)^{-1} \quad \mbox{with}  \\
&{}& h_2 \quad \mbox{defined by} \quad h_2 = - \frac{5}{6}   T p + \frac{\rho T}{2}
\frac{\partial p }{\partial \rho} + \frac{T^2}{2 \rho} \frac{\left( \frac{\partial p
}{\partial T} \right)^{2}}{\frac{\partial \varepsilon}{\partial T}} \, . \nonumber
\end{eqnarray}
Moreover, we have:
\begin{eqnarray}\label{v.1}
&{}& \bullet  \quad   (h_{1,0,1})_{Eq.} = \frac{\partial h_{0,1,0}}{\partial \lambda} =
-2T (p+ \rho \varepsilon) \, , \quad  \quad  \quad  \quad \quad  \quad  \quad  \quad
\quad \quad  \quad  \quad \quad  \quad  \quad  \quad \quad  \quad  \quad  \quad \quad
\quad \quad  \quad
\end{eqnarray}
where we have used (\ref{cris11}) in the first passage and $(\ref{cris28})_2$ in the
second passage,
\begin{eqnarray}\label{v.2}
&{}& \bullet  \quad   (h_{2,0,0})_{Eq.} = -  \frac{1}{2 \lambda} \frac{\partial
h_{0,0,0}}{\partial \mu} = -T \rho  \, , \quad  \quad  \quad  \quad \quad \quad \quad
\quad \quad \quad  \quad  \quad \quad  \quad  \quad  \quad \quad \quad \quad \quad \quad
\quad \quad  \quad \quad
\end{eqnarray}
where we have used (\ref{cris11}) in the first passage and $(\ref{cris13})_1$ in the
second passage,
\begin{eqnarray}\label{v.2bis}
&{}& \bullet  \quad   (\phi_{1,1,0})_{Eq.} = 3T^2 h_{0,0,0} =  - 3 p T  \, , \quad \quad
\quad  \quad \quad \quad \quad \quad \quad \quad  \quad  \quad \quad \quad \quad \quad
\quad \quad \quad \quad \quad \quad \quad  \quad \quad  \quad
\end{eqnarray}
where we have used $(\ref{cris12})_3$ and $(\ref{cris11})_1$ in the first passage and
(\ref{cris17}) in the second passage,
\begin{eqnarray}\label{v.3}
&{}& \bullet  \quad   (\phi_{1,0,0})_{Eq.} = p  \, , \quad \quad \quad  \quad \quad \quad
\quad \quad \quad \quad  \quad  \quad \quad \quad \quad \quad \quad \quad \quad \quad
\quad \quad \quad  \quad \quad  \quad \quad \quad \quad \quad \quad \quad \quad
\end{eqnarray}
where we have used $(\ref{cris12})_1$, $(\ref{cris11})_1$, (\ref{cris17}),
\begin{eqnarray}\label{v.3bis}
&{}& \bullet  \quad   \frac{\partial \phi_{1,0,0}}{\partial \mu} = \frac{\partial^3
}{\partial \mu^3} \left( -  \frac{1}{2 \lambda} \psi_1 \right) =- T \rho  \, , \quad
\quad \quad  \quad \quad \quad \quad \quad \quad \quad  \quad  \quad \quad  \quad \quad
\quad \quad \quad \quad \quad \quad \quad \quad  \quad \quad
\end{eqnarray}
where we have used $(\ref{cris12})_1$ and $(\ref{p.2q})_1$,
\begin{eqnarray}\label{v.3ter}
&{}& \bullet  \quad   \frac{\partial \phi_{1,0,0}}{\partial \lambda} = \frac{\partial
}{\partial \lambda} \left( -  \frac{1}{2 \lambda} h_{0,0,0} \right) =  -2 \rho T \left(
\varepsilon + \frac{p}{\rho} \right) \, , \quad \quad \quad  \quad \quad \quad \quad
\quad \quad \quad  \quad  \quad \quad  \quad \quad \quad \quad \quad \quad
\end{eqnarray}
where we have used $(\ref{cris12})_1$ and $(\ref{cris11})_1$ in the first passage and a
part of $(\ref{cris28})_2$ in the second passage.  \\
Now, from $(\ref{cris11})_6$ and $(\ref{cris12})_2$ it follows $h_{0,0,2}=\frac{\partial
}{\partial \lambda}\phi_{0,0,1}$. By using this result, we may find
\begin{eqnarray}\label{z.1}
\bullet  \quad h_{0,0,2} = \frac{\partial \phi_{0,0,1}}{\partial \rho} \frac{\partial
\rho}{\partial \lambda} + \frac{\partial \phi_{0,0,1}}{\partial T} \frac{\partial
T}{\partial \lambda} \stackrel{(1)}{=} \frac{\partial \phi_{0,0,1}}{\partial \rho}  2T
\left( T \frac{\partial p}{\partial T} - p - \rho \varepsilon \right)\left(
\frac{\partial p}{\partial \rho}
\right)^{-1} - 2 \frac{\partial \phi_{0,0,1}}{\partial T} T^2 = \\
\stackrel{(2)}{=} 4 \left( \varepsilon + \frac{p}{\rho} \right) T \left( T \frac{\partial
p}{\partial T} - p - \rho \varepsilon \right) - 2 T^2 \frac{\partial
\phi_{0,0,1}}{\partial T} \nonumber
\end{eqnarray}
where in the passage denoted with $\stackrel{(1)}{=}$ we have used $(\ref{cris13})_1$,
$(\ref{cris11})_1$ and $(\ref{cris21})_2$; moreover, in the passage denoted with
$\stackrel{(2)}{=}$ we have used eq. (\ref{cris25}). So, up to now, for the first order
we have found every thing in terms of $p(\rho,T)$, $\varepsilon(\rho,T)$,
$\phi_{0,0,1}(\rho,T)$ and $\phi_{0,1,1}$. On the last two of these functions we have the
condition (\ref{cris25}) and the following ones
\begin{eqnarray}\label{z.2}
\bullet  \quad \frac{\partial \phi_{0,1,1}}{\partial \rho}= \frac{\partial
\phi_{0,1,1}}{\partial \lambda} \frac{\partial \lambda}{\partial \rho} + \frac{\partial
\phi_{0,1,1}}{\partial \mu} \frac{\partial \mu}{\partial \rho} \stackrel{(1)}{=}
 3 \frac{\partial^3 }{\partial \lambda \partial \mu^2} \left[ \left( \frac{1}{2\lambda}\right)^2  \psi_1
 \right] \frac{-1}{\rho T} \frac{\partial p}{\partial \rho} =
 \quad \quad \quad  \quad \quad \quad \quad \quad  \quad \quad\\
=  3 \frac{\partial }{\partial \lambda} \left[ \left( \frac{1}{2\lambda} \right)^2
h_{0,0,0} \right] \frac{-1}{\rho T} \frac{\partial p}{\partial \rho} = \frac{3}{2} \left(
\frac{-1}{\lambda^3}  h_{0,0,0}+ \frac{1}{2 \lambda^2}  \frac{\partial
h_{0,0,0}}{\partial \lambda} \right) \frac{-1}{\rho T} \frac{\partial p}{\partial \rho}
\stackrel{(2)}{=} - 6 \left( \varepsilon + 2 \frac{p}{\rho} \right) T \frac{\partial
p}{\partial \rho} \, , \nonumber
\end{eqnarray}
where in the passage denoted with $\stackrel{(1)}{=}$ we have used $(\ref{cris12})_4$,
and $(\ref{cris24})_1$; moreover, in the passage denoted with $\stackrel{(2)}{=}$ we have
used eq. (\ref{cris17}) and $(\ref{cris13})_3$.
\begin{eqnarray}\label{z.3}
\bullet  \, \frac{\partial \phi_{0,0,1}}{\partial T}= \frac{\partial
\phi_{0,0,1}}{\partial \lambda} \frac{\partial \lambda}{\partial T} + \frac{\partial
\phi_{0,0,1}}{\partial \mu} \frac{\partial \mu}{\partial T} \stackrel{(1)}{=}
 \frac{\partial^3 }{\partial \lambda \partial \mu^2}\left( \frac{-1}{2\lambda}  \psi_1
 \right) \frac{-1}{\rho T^2} \left( T \frac{\partial p}{\partial T} - p - \rho \varepsilon
 \right)+  \frac{-1}{2T^2} \frac{\partial}{\partial \lambda} H_{1,0,1,0} + \\
 + \frac{\partial^3 }{\partial \lambda^2 \partial \mu}\left( \frac{-1}{2\lambda}  \psi_1
 \right) \frac{-1}{2T^2} =  \frac{\partial }{\partial \lambda} \left( \frac{1}{2\lambda} h_{0,0,0} \right) \frac{1}{\rho T^2}
 \left( T \frac{\partial p}{\partial T} - p - \rho \varepsilon \right) +
 \frac{-1}{2T^2} \frac{\partial}{\partial \lambda} H_{1,0,1,0}+   \nonumber \\
+ \frac{1}{2T^2} \frac{\partial^3 }{\partial \lambda^2 \partial \mu}\left(
\frac{1}{2\lambda}  \psi_1
 \right) = \left(
\frac{-1}{2 \lambda^2}  h_{0,0,0}+ \frac{1}{2 \lambda}  \frac{\partial
h_{0,0,0}}{\partial \lambda} \right) \frac{1}{\rho T^2}
 \left( T \frac{\partial p}{\partial T} - p - \rho \varepsilon \right)+
 \frac{-1}{2T^2} \frac{\partial}{\partial \lambda} H_{1,0,1,0}+  \nonumber\\
+ \frac{1}{2T^2}
 \frac{\partial^3 }{\partial \lambda^2 \partial \mu}\left( \frac{1}{2\lambda}  \psi_1
 \right) \stackrel{(2)}{=} (2p+2 \rho \varepsilon ) \frac{1}{\rho T}
 \left( T \frac{\partial p}{\partial T} - p - \rho \varepsilon \right) +
 \frac{-1}{2T^2} \frac{\partial}{\partial \lambda} H_{1,0,1,0}+  \frac{1}{2T^2}
 \frac{\partial^3 }{\partial \lambda^2 \partial \mu}\left( \frac{1}{2\lambda}  \psi_1
 \right)
 \, , \nonumber
\end{eqnarray}
where in the passage denoted with $\stackrel{(1)}{=}$ we have used $(\ref{cris12})_2$ and
$(\ref{cris24})$; moreover, in the passage denoted with $\stackrel{(2)}{=}$ we have used
eq. (\ref{cris17}) and $(\ref{cris13})_3$.
\begin{eqnarray}\label{aa.1}
\bullet  \quad \frac{\partial \phi_{0,1,1}}{\partial T}= \frac{\partial
\phi_{0,1,1}}{\partial \lambda} \frac{\partial \lambda}{\partial T} + \frac{\partial
\phi_{0,1,1}}{\partial \mu} \frac{\partial \mu}{\partial T} \stackrel{(1)}{=} 3
 \frac{\partial^3 }{\partial \lambda \partial \mu^2} \left[ \left( \frac{1}{2\lambda} \right)^2  \psi_1
 \right] \frac{-1}{\rho T^2} \left( T \frac{\partial p}{\partial T} - p - \rho \varepsilon
 \right)+  \quad \quad \quad \quad \quad \\
 + \left\{ 3 \frac{\partial^3 }{\partial \lambda^2 \partial \mu} \left[ \left( \frac{1}{2\lambda} \right)^2
 \psi_1 \right]+ \frac{\partial
H_{1,1,1,0}}{\partial \lambda} \right\} \frac{-1}{2 T^2} = 3 \frac{\partial }{\partial
\lambda} \left( \frac{1}{4 \lambda^2} h_{0,0,0} \right) \frac{-1}{\rho T^2}
 \left( T \frac{\partial p}{\partial T} - p - \rho \varepsilon \right) + \nonumber \\
 -  \frac{3}{2T^2}
 \frac{\partial^3 }{\partial \lambda^2 \partial \mu}\left( \frac{1}{4 \lambda^2}  \psi_1
 \right) - \frac{1}{2 T^2} \frac{\partial
H_{1,1,1,0}}{\partial \lambda} = \nonumber \\
 = \frac{3}{4} \left(
\frac{-2}{\lambda^3}  h_{0,0,0}+ \frac{1}{\lambda^2}  \frac{\partial h_{0,0,0}}{\partial
\lambda} \right) \frac{-1}{\rho T^2}
 \left( T \frac{\partial p}{\partial T} - p - \rho \varepsilon \right) -  \frac{3}{2T^2}
 \frac{\partial^3 }{\partial \lambda^2 \partial \mu}\left( \frac{1}{4\lambda^2}  \psi_1
 \right)- \frac{1}{2 T^2} \frac{\partial
H_{1,1,1,0}}{\partial \lambda} = \nonumber\\
\stackrel{(2)}{=} \frac{3}{4} ( 16 p T^2 + 8 T^2 \rho \varepsilon ) \frac{-1}{\rho T^2}
\left( T \frac{\partial p}{\partial T} - p - \rho \varepsilon \right) -  \frac{3}{2T^2}
 \frac{\partial^3 }{\partial \lambda^2 \partial \mu}\left( \frac{1}{4\lambda^2}  \psi_1
 \right) - \frac{1}{2 T^2} \frac{\partial
H_{1,1,1,0}}{\partial \lambda}
 \, , \nonumber
\end{eqnarray}
where in the passage denoted with $\stackrel{(1)}{=}$ we have used $(\ref{cris12})_4$ and
$(\ref{cris24})$; moreover, in the passage denoted with $\stackrel{(2)}{=}$ we have used
eq. (\ref{cris17}) and $(\ref{cris13})_3$.  \\
Now we note that $(\ref{cris12})_{2,4}$, $(\ref{z.3})$ and $(\ref{aa.1})$ are 4 equations
in the 3 unknowns  $\frac{\partial \psi_1}{\partial \mu}$, $\frac{\partial^2
\psi_1}{\partial \lambda \partial \mu}$,  $\frac{\partial^3 \psi_1}{\partial \lambda^2
\partial \mu}$; it follows that the determinant of the complete matrix must be zero, that
is
\begin{eqnarray*}
\left| \begin{array}{ccccccc}
  \frac{1}{2 \lambda^2} &  & - \frac{1}{2 \lambda} &  & 0 &  & - \phi_{0,0,1} + H_{1,0,1,0} \\
   &  &  &  &  &  &  \\
 -\frac{3}{2} \frac{1}{\lambda^3}   &  & \frac{3}{4 \lambda^2}  &  & 0 &  & - \phi_{0,1,1} + H_{1,1,1,0}\\
   &  & & & &  &  \\
  \frac{1}{2 \lambda^3 T^2}  &  & \frac{-1}{2 \lambda^2 T^2} &  & \frac{1}{4 \lambda T^2}  &  &
  - \frac{\partial \phi_{0,0,1}}{\partial T} + 2 \left(\frac{p}{\rho}+ \varepsilon \right) \frac{1}{T}
 \left( T \frac{\partial p}{\partial T} - p - \rho \varepsilon \right) +
 \frac{-1}{2T^2} \frac{\partial}{\partial \lambda} H_{1,0,1,0}\\
   &  &  &  & &  &  \\
 \frac{-9}{4 \lambda^4T^2}  &  & \frac{3}{2 \lambda^3 T^2}  &  & \frac{-3}{8 \lambda^2 T^2}  &  &
  - \frac{\partial \phi_{0,1,1}}{\partial T} - 6 ( \frac{2 p }{\rho} +  \varepsilon )
\left( T \frac{\partial p}{\partial T} - p - \rho \varepsilon \right) - \frac{1}{2 T^2}
\frac{\partial H_{1,1,1,0}}{\partial \lambda}
\end{array} \right| =0 \, .
\end{eqnarray*}
Now this determinant can be simplified by adding to its first column the second one
multiplied by $\frac{2}{\lambda}$ and the third one multiplied by $\frac{2}{\lambda^2}$;
moreover, in line 3, column 4 we can substitute $\frac{\partial}{\partial \lambda}
H_{1,0,1,0}$ from $(\ref{cris12a})_1$, so that it becomes
\begin{eqnarray*}
\left| \begin{array}{ccccccc}
  -\frac{1}{2 \lambda^2} &  & - \frac{1}{2 \lambda} &  & 0 &  & - \phi_{0,0,1} + H_{1,0,1,0} \\
   &  &  &  &  &  &  \\
0   &  & \frac{3}{4 \lambda^2}  &  & 0 &  & - \phi_{0,1,1} + H_{1,1,1,0}\\
   &  & & & &  &  \\
0 &  & \frac{-1}{2 \lambda^2 T^2} &  & \frac{1}{4 \lambda T^2}  &  &
  - \frac{\partial \phi_{0,0,1}}{\partial T} + 2 \left(\frac{p}{\rho}+ \varepsilon \right) \frac{1}{T}
 \left( T \frac{\partial p}{\partial T} - p - \rho \varepsilon \right)+ \frac{2}{3T}
 \frac{\partial}{\partial \lambda} \left( \lambda^2 H_{1,1,1,0} \right) \\
   &  &  &  & &  &  \\
0  &  & \frac{3}{2 \lambda^3 T^2}  &  & \frac{-3}{8 \lambda^2 T^2}  &  &
  - \frac{\partial \phi_{0,1,1}}{\partial T} - 6 ( \frac{2 p }{\rho} +  \varepsilon )
\left( T \frac{\partial p}{\partial T} - p - \rho \varepsilon \right) - \frac{1}{2 T^2}
\frac{\partial H_{1,1,1,0}}{\partial \lambda}
\end{array} \right| =0 \, .
\end{eqnarray*}
(It is interesting that now $H_{1,1,1,0}$ disappears and the result depends on
$\phi_{0,0,1}$ only through $\frac{\partial \phi_{0,0,1}}{\partial T}$). The resulting
expression can be used to find
\begin{eqnarray}\label{ab.1}
\bullet  \quad \frac{\partial \phi_{0,1,1}}{\partial T} = \frac{2}{T} \phi_{0,1,1} - 3T
\frac{\partial \phi_{0,0,1}}{\partial T} - 6 \frac{p }{\rho} \left( T \frac{\partial
p}{\partial T} - p - \rho \varepsilon \right)
\end{eqnarray}
This result can be rewritten by inserting $\frac{\partial \phi_{0,0,1}}{\partial T}$ from
(\ref{cris27}), so that it becomes
\begin{eqnarray}\label{alfa.5}
\bullet  \quad \frac{\partial \phi_{0,1,1}}{\partial T} = \frac{2}{T} \phi_{0,1,1}  - 6T
\left(\frac{2p}{\rho}+ \varepsilon \right) \frac{\partial p}{\partial T} +3 \frac{h_4}{T}
+ 6 p \left(\frac{p}{\rho}+ \varepsilon \right) \, .
\end{eqnarray}
After that, from (\ref{z.1}) with (\ref{cris27}) we obtain
\begin{eqnarray}\label{beta.1}
\bullet  \quad h_{0,0,2} = 2 h_4 - 4 \rho T \left(\frac{p}{\rho}+ \varepsilon \right)^2
\, .
\end{eqnarray}
Moreover, it will be useful for the sequel to know the determinant
\begin{eqnarray}\label{beta.22}
\bullet  \quad D_1 = \left| \begin{array}{ccc}
  (h_{2,0,0})_{Eq.} &  &  (h_{1,0,1})_{Eq.}  \\
   &  &  \\
   (h_{1,0,0})_{Eq.}  &  &  (h_{0,0,2})_{Eq.}
\end{array} \right| = -2h_4 \rho T \, ,
\end{eqnarray}
where in the second passage we have used (\ref{v.2}), (\ref{v.1}) and (\ref{beta.1}). \\
To conclude this preparatory work, let us note that
\begin{eqnarray}\label{gamma.1}
\bullet \quad \quad \frac{\partial \phi_{0,0,1}}{\partial \mu} \stackrel{(1)}{=}
\frac{\partial^3}{\partial \lambda \partial \mu^2} \left( \frac{-1}{2\lambda}  \psi_1
 \right) \stackrel{(2)}{=} \frac{\partial}{\partial \lambda} \left( \frac{-1}{2\lambda} h_{0,0,0}
 \right)= \frac{1}{2\lambda^2} h_{0,0,0} - \frac{1}{2\lambda} \frac{\partial h_{0,0,0}}{\partial
 \lambda}= \\
\stackrel{(3)}{=} - 2 p T -2 \rho \varepsilon T = - 2 \rho T \left(\frac{p}{\rho}+
\varepsilon \right) \, , \nonumber
\end{eqnarray}
where in the passage denoted with $\stackrel{(1)}{=}$ we have used $(\ref{cris12})_2$, in
the passage denoted with $\stackrel{(2)}{=}$ we have used eq. $(\ref{cris11})_1$ and in
the passage denoted with $\stackrel{(3)}{=}$ we have used eq. $(\ref{cris17})$ and
$(\ref{cris13})_3$.  \\
\begin{eqnarray*}
\frac{\partial \phi_{0,0,1}}{\partial T} =\frac{\partial \phi_{0,0,1} }{\partial \lambda}
\frac{-1}{2T^2}  + \frac{\partial \phi_{0,0,1} }{\partial \mu} \frac{\partial \mu
}{\partial T} \, ,
\end{eqnarray*}
or, by using $(\ref{cris27})$, $(\ref{gamma.1})$ and $(\ref{cris24})$,
\begin{eqnarray*}
2 \left(\frac{p}{\rho}+ \varepsilon \right) \frac{\partial p}{\partial T} -
\frac{h_4}{T^2}= \frac{\partial \phi_{0,0,1} }{\partial \lambda} \frac{-1}{2T^2} + 2 \rho
T \left(\frac{p}{\rho}+ \varepsilon \right) \frac{1}{\rho T^2} \left( T \frac{\partial
p}{\partial T} - p - \rho \varepsilon \right) \, ,
\end{eqnarray*}
from which we deduce
\begin{eqnarray}\label{gamma.2}
\bullet \quad \quad \frac{\partial \phi_{0,0,1}}{\partial \lambda}= 2 h_4 - 4 \rho T
\left(\frac{p}{\rho}+ \varepsilon \right)^2 \, .
\end{eqnarray}
\subsection{The variables at first order with respect to equilibrium.}
Let us substitute (\ref{cris9}) and (\ref{cris10}) in $(\ref{2bis.3})_1$ and calculate
the result at first order with respect to Equilibrium and for $v^i=0$. So we obtain
\begin{eqnarray}\label{S.2}
&{}& 0= \left( \frac{\partial^2 h_{0,0,0}}{\partial \mu^2} \right)_{Eq.}(\mu -
\mu_{Eq.})^{(1)}+ \left( \frac{\partial^2 h_{0,0,0}}{\partial \lambda \partial
\mu}\right)_{Eq.} \left(\lambda - \frac{1}{2T} \right)^{(1)}+ \left( \frac{\partial
h_{0,1,0}}{\partial \mu}\right)_{Eq.} (\mu_{ll})^{(1)} \, , \\
&{}& 0= \left( h_{2,0,0} \right)_{Eq.}(\mu^i)_{(1)}+ \left( h_{1,0,1}\right)_{Eq.}
\left(\lambda^i\right)_{(1)} \, , \nonumber \\
&{}& \pi \delta^{ij} + \hat{F}^{ij} = \left[ \left( \frac{\partial h_{0,1,0}}{\partial
\mu} \right)_{Eq.}(\mu - \mu_{Eq.})^{(1)}+ \left( \frac{\partial h_{0,1,0}}{\partial
\lambda}\right)_{Eq.} \left(\lambda - \frac{1}{2T} \right)^{(1)}  \right]\delta^{ij} + \nonumber \\
&{}& \quad \quad \quad \quad \quad \quad + \left( h_{0,2,0}\right)_{Eq.} \left[
\frac{1}{3}(\mu_{ll})^{(1)} \delta^{ij} + \frac{2}{3} \mu^{ij}_{(1)} \right] \, , \nonumber \\
&{}& 0= \left( \frac{\partial^2 h_{0,0,0}}{\partial \lambda \partial \mu}
\right)_{Eq.}(\mu - \mu_{Eq.})^{(1)}+ \left( \frac{\partial^2 h_{0,0,0}}{\partial
\lambda^2 }\right)_{Eq.} \left(\lambda - \frac{1}{2T} \right)^{(1)}+ \left(
\frac{\partial h_{0,1,0}}{\partial \lambda}\right)_{Eq.} (\mu_{ll})^{(1)} \, , \nonumber \\
&{}& 2 q^i = \hat{G}^{ill}= \left( h_{1,0,1} \right)_{Eq.}(\mu^i)_{(1)}+ \left(
h_{0,0,2}\right)_{Eq.} \left(\lambda^i\right)_{(1)} \, . \nonumber
\end{eqnarray}
Well,  $(\ref{S.2})_{1,4}$ and the trace of $(\ref{S.2})_{3}$ constitute a system  from
which we can obtain
\begin{eqnarray}\label{T.1}
 (\mu -\mu_{Eq.})^{(1)} = \frac{\pi}{D} \left| \begin{array}{ccc}
   \left( \frac{\partial^2 h_{0,0,0}}{\partial \lambda \partial \mu}
\right)_{Eq.} &  & - \rho T \\
    &  & \\
   \left( \frac{\partial^2 h_{0,0,0}}{\partial \lambda^2}
\right)_{Eq.} &  & -2T (p+\varepsilon \rho )
 \end{array} \right| \, , \\
\nonumber \\
  \left(\lambda - \frac{1}{2T} \right)^{(1)} = - \frac{\pi}{D} \left| \begin{array}{ccc}
   \left( \frac{\partial^2 h_{0,0,0}}{ \partial \mu^2}
\right)_{Eq.} &  & - \rho T \\
    &  & \\
   \left( \frac{\partial^2 h_{0,0,0}}{\partial \lambda \partial \mu}
\right)_{Eq.} &  & -2T (p+\varepsilon \rho )
 \end{array} \right| \, , \nonumber\\
 \nonumber\\
  (\mu_{ll})^{(1)}= \frac{\pi}{D} \left| \begin{array}{ccc}
   \left( \frac{\partial^2 h_{0,0,0}}{\partial \mu^2}
\right)_{Eq.} &  & \left( \frac{\partial^2 h_{0,0,0}}{\partial \lambda \partial \mu}
\right)_{Eq.}  \\
    &  & \\
   \left( \frac{\partial^2 h_{0,0,0}}{\partial \lambda \partial \mu}
\right)_{Eq.}  &  & \left( \frac{\partial^2 h_{0,0,0}}{\partial \lambda^2} \right)_{Eq.}
 \end{array} \right| \, , \nonumber
\end{eqnarray}
with $D$ given by (\ref{cris31}) and where we have taken into account (\ref{cris28}).  \\
Moreover, the traceless part of $(\ref{S.2})_3$ is
\begin{eqnarray}\label{ad.2}
\mu^{<ij>}_{(1)}= \frac{3}{2} \left[ (h_{0,2,0})_{Eq.} \right]^{-1} \hat{F}^{<ij>} =
-\frac{1}{2pT} \hat{F}^{<ij>} \, ,
\end{eqnarray}
while $(\ref{S.2})_{2,5}$ give
\begin{eqnarray}\label{ad.1}
(\mu^i)_{(1)}= 2 q^i \quad \frac{-(h_{1,0,1})_{Eq.}}{D_1} \quad , \quad
\left(\lambda^i\right)_{(1)}  = 2 q^i \quad \frac{(h_{2,0,0})_{Eq.}}{D_1} \, ,
\end{eqnarray}
with $D_1$ given by (\ref{beta.22}).  \\
After that, let us substitute (\ref{cris9}) and (\ref{cris10}) in $(\ref{2bis.3})_2$ and
calculate the result at first order with respect to Equilibrium and for $v^i=0$. So we
obtain
\begin{eqnarray}\label{u.1}
\hat{F}^{kij}_{(1)}= \delta ^{(ki} q^{j)}  \left[ 2 (\phi_{1,1,0})_{Eq.} \quad
\frac{-(h_{1,0,1})_{Eq.}}{D_1} + 2 (\phi_{0,1,1})_{Eq.} \quad
\frac{(h_{2,0,0})_{Eq.}}{D_1} \right] \, ,  \\
\hat{G}^{kill}_{(1)}=   \left[ \left( \frac{\partial \phi_{0,0,1}}{\partial
\partial \mu}\right)_{Eq.}  (\mu -\mu_{Eq.})^{(1)}
+ \left( \frac{\partial \phi_{0,0,1}}{\partial \lambda}\right)_{Eq.}   \left(\lambda -
\frac{1}{2T} \right)^{(1)} \right] \delta ^{ki}
+ \nonumber \\
+ (\phi_{0,1,1})_{Eq.} \left( \frac{5}{9} (\mu_{ll})^{(1)} \delta ^{ki} + \frac{2}{3}
\mu^{<ki>}_{(1)} \right) \, .  \nonumber
\end{eqnarray}
Now, thanks to (\ref{cris21}), eqs. (\ref{T.1}) become
\begin{eqnarray}\label{ae.1}
(\mu -\mu_{Eq.})^{(1)} = \frac{\pi}{D} \left| \begin{array}{ccc}
   2 T \left[ T \left(\frac{\partial p}{\partial T}\right)_{\rho} -p - \rho \varepsilon \right]
\left(\frac{\partial p}{\partial \rho}\right)_{T}^{-1}&  & - \rho T \\
    &  & \\
   - 4 \rho T^2 \left(\frac{\partial \varepsilon}{\partial T}\right)_{\rho} - \frac{4T}{\rho}
\left[ T \left(\frac{\partial p}{\partial T}\right)_{\rho} -p - \rho \varepsilon
\right]^2 \left(\frac{\partial p}{\partial \rho}\right)_{T}^{-1} &  & -2T (p+\varepsilon
\rho )
 \end{array} \right| =  \\
 \nonumber \\
= \left\{ -4 \rho^2 T^3 \left(\frac{\partial \varepsilon}{\partial T}\right)_{\rho} - 4
T^3 \left[ T \left(\frac{\partial p}{\partial T}\right)_{\rho} -p - \rho \varepsilon
\right] \left(\frac{\partial p}{\partial T}\right)_{\rho} \left(\frac{\partial
p}{\partial \rho}\right)_{T}^{-1} \right\} \frac{\pi}{D} \,
, \nonumber \\
 \nonumber \\
\left(\lambda - \frac{1}{2T} \right)^{(1)} = - \frac{\pi}{D} \left| \begin{array}{ccc}
   - \rho T
\left(\frac{\partial p}{\partial \rho}\right)_{T}^{-1} &  & - \rho T \\
    &  & \\
   2 T \left[ T \left(\frac{\partial p}{\partial T}\right)_{\rho} -p - \rho \varepsilon \right]
\left(\frac{\partial p}{\partial \rho}\right)_{T}^{-1} &  & -2T (p+\varepsilon \rho )
 \end{array} \right| = \nonumber \quad \quad \quad \quad \quad \quad \\
 \nonumber \\
 = -2 \rho T^3 \left(\frac{\partial p}{\partial T}\right)_{\rho}
 \left(\frac{\partial p}{\partial \rho}\right)_{T}^{-1} \frac{\pi}{D} \, , \nonumber\\
 \nonumber\\
  (\mu_{ll})^{(1)}= 4 \rho^2 T^3 \left(\frac{\partial \varepsilon}{\partial T}\right)_{\rho}
\left(\frac{\partial p}{\partial \rho}\right)_{T}^{-1}
  \frac{\pi}{D} \, . \nonumber  \quad \quad \quad \quad \quad \quad  \quad \quad \quad \quad \quad
  \quad \quad \quad \quad \quad \quad \quad  \quad \quad \quad \quad
\end{eqnarray}
Instead of this, eqs. (\ref{ad.1}), thanks to (\ref{beta.22}), (\ref{v.1}), (\ref{v.2}),
become
\begin{eqnarray}\label{af.1}
(\mu^i)_{(1)}= -  \frac{2}{h_4} \left(\frac{p}{\rho}+ \varepsilon \right) q^i \quad ,
\quad \left(\lambda^i\right)_{(1)}  = \frac{1}{h_4} \, q^i \, .
\end{eqnarray}
Thanks to this result and also to $(\ref{v.2})_2$, (\ref{v.1}) and (\ref{beta.22}), the
closure $(\ref{u.1})_1$ becomes
\begin{eqnarray}\label{af.2}
\hat{F}^{kij}_{(1)}= \frac{1}{h_4} \left[ \phi_{0,1,1} + 6 p T \left(\frac{p}{\rho}+
\varepsilon \right) \right] \delta ^{(ki} q^{j)} \, .
\end{eqnarray}
Similarly, the closure $(\ref{u.1})_2$, thanks also to (\ref{gamma.1}), (\ref{gamma.2}),
(\ref{ad.2}), (\ref{cris31}) becomes
\begin{eqnarray}\label{af.3}
\hat{G}^{kill}_{(1)}=   \left[ \frac{h_4}{2h_2} \left( \frac{5}{6} K -
\frac{\frac{\partial p}{\partial T}}{\rho \frac{\partial \varepsilon}{\partial T}}
\right) + 2 \left(\frac{p}{\rho}+ \varepsilon \right) \right] \pi \, \delta^{ki} + \left[
- \frac{h_4}{2} \frac{K}{p T} + 2 \left(\frac{p}{\rho}+ \varepsilon \right)
\right] \hat{F}^{<ki>} \, ,   \\
\mbox{with} \quad K= \frac{2}{3} \frac{1}{h_4} \left[ \phi_{0,1,1} + 6 p T
\left(\frac{p}{\rho}+ \varepsilon \right) \right] \, .  \nonumber
\end{eqnarray}

\subsection{The entropy density and its flux, up to second order.}
From properties of homogeneous functions, we have that the expression of first order of
the entropy density is equal to $h^{(1)}= \pi^A \frac{\partial h^{(1)} }{\partial \pi^A}$
where $\pi^A$ denotes $\pi$, $\hat{F}^{<ij>}$, $q^i$. Moreover, from $(\ref{2bis.1})_1$
it follows
\begin{eqnarray*}
&{}&   \mu_{<ij>} = \frac{\partial h}{\partial \hat{F}^{<ij>}} \quad , \quad
 \mu_{ll} = \frac{\partial h}{\partial \pi} \quad , \quad
 \lambda_i= \frac{1}{2} \frac{\partial h}{\partial q^i} \quad  .
\end{eqnarray*}
It follows that $h^{(1)}= \pi^A \left( \frac{\partial h }{\partial \pi^A} \right)^{(0)}
=0$, that is the homogeneous part of the entropy density at first order is zero. By
proceeding in a similar way for the entropy density at second order, we obtain
\begin{eqnarray}\label{ag.1}
h^{(2)} = \frac{1}{2} \frac{\partial^2 h^{(2)} }{\partial \pi^A \partial \pi^B} \pi^A
\pi^B = \frac{1}{2} \left( \frac{\partial h }{\partial \pi^A} \right)^{(1)} \pi^A =
\frac{1}{2} (\mu_{ll})^{(1)} \pi + \frac{1}{2}  (\mu_{<ij>})^{(1)} \hat{F}^{<ij>} +
(\lambda_{i})^{(1)} q^i =
\end{eqnarray}
\begin{eqnarray*}
&{}& \stackrel{1}{=} 2 \rho^2 T^3\frac{\partial \varepsilon}{\partial T}
\left(\frac{\partial p}{\partial \rho}\right)^{-1} \frac{\pi^2}{D} - \frac{1}{4pT}
\hat{F}^{<ij>} \hat{F}_{<ij>} + \frac{1}{h_4} q^i q_i   \stackrel{2}{=} \frac{1}{4 h_2}
\pi^2 +  \frac{1}{4 h_3}  \hat{F}^{<ij>} \hat{F}_{<ij>} + \frac{1}{h_4} q^i q_i \quad  ,
\end{eqnarray*}
where in the passage denoted with $\stackrel{1}{=}$ we have used $(\ref{ae.1})_3$,
$(\ref{ad.2})$, $(\ref{af.1})$, while in the passage denoted with $\stackrel{2}{=}$ we
have used (\ref{cris31}); moreover, we have used the new function
\begin{eqnarray}\label{ag.2}
h_3 = - pT \, .
\end{eqnarray}
We appreciate that this result is exactly the same of eq. (55) in \cite{1} with $h_3$
given in their eq. $(43)_2$. Also the expression for $h_2$ given in eq. $(43)_1$ of
\cite{1} corresponds to the present $(\ref{cris31})_2$. \\
For what concerns the entropy flux $h^k$, we have that $(\ref{2bis.1})_2$ implies
\begin{eqnarray}\label{ah.1}
&{}& \frac{\partial h^k}{\partial \pi} = \mu^k + \mu_{ij} \frac{\partial
F^{kij}}{\partial \pi} + \lambda_i \frac{\partial G^{kill}}{\partial \pi} \, , \\
&{}& \frac{\partial h^k}{\partial F_{<ab>}} = \mu^{<a} \delta^{b>k} + \mu_{ij}
\frac{\partial
F^{kij}}{\partial F_{<ab>}} + \lambda_i \frac{\partial G^{kill}}{\partial F_{<ab>}} \, , \nonumber \\
&{}& \frac{\partial h^k}{\partial q_a} = \mu_{ij} \frac{\partial F^{kij}}{\partial q_a} +
2 \lambda \delta^{ka} + \lambda_i \frac{\partial G^{kill}}{\partial q_a} \, , \nonumber
\end{eqnarray}
which can be calculated in $v^j=0$ so transforming themselves in the corresponding ones
denoted with an $ \hat{}$. \\
By calculating them at equilibrium, the first two will say that $\hat{h}^k_{(1)}$ doesn'
t depend on $\pi$, nor on $F_{<ab>}$, while the third one becomes
\begin{eqnarray}\label{ah.2}
&{}& \frac{\partial \hat{h}^k_{(1)}}{\partial q_a} =  \frac{1}{T} \delta^{ka} \quad
\mbox{from which we obtain the well known expression} \quad  \hat{h}^k_{(1)} =
\frac{q^k}{T} \, .
\end{eqnarray}
We note now that, by substituting $\phi_{0,1,1}$ from $(\ref{af.3})_2$ in (\ref{af.2}),
we obtain
\begin{eqnarray}\label{ai.2}
&{}& \hat{F}^{kij}_{(1)} =  \frac{3}{2} \, K \, \delta^{(ij} q^{k)} \, .
\end{eqnarray}
By calculating (\ref{ah.1}) at first order with respect to equilibrium, thanks also to
eq. (\ref{ai.2}), (\ref{af.1}), (\ref{ad.2}), (\ref{ae.1}), (\ref{af.2}),  they become
\begin{eqnarray*}
&{}& \frac{\partial \hat{h}^k_{(2)}}{\partial \pi} = \frac{1}{2h_2} \left( \frac{5}{6} K
- \frac{\frac{\partial p}{\partial T}}{\rho \frac{\partial \varepsilon}{\partial T}}
\right)
q^k \, , \\
&{}& \frac{\partial \hat{h}^k_{(2)}}{\partial F_{<ab>}} = - \frac{1}{2pT} K \delta^{k<a}
q^{b>}  \, ,  \\
&{}& \frac{\partial \hat{h}^k_{(2)}}{\partial q_a} = \frac{1}{2h_2} \left( \frac{5}{6} K
- \frac{\frac{\partial p}{\partial T}}{\rho \frac{\partial \varepsilon}{\partial T}}
\right) \pi \delta^{ka}  - \frac{1}{2pT} K \hat{F}^{<ka>} \, ,
\end{eqnarray*}
from which it follow
\begin{eqnarray}\label{ai.1}
&{}& \hat{h}^k_{(2)} =  \pi q^k  \frac{1}{2h_2} \left( \frac{5}{6} K -
\frac{\frac{\partial p}{\partial T}}{\rho \frac{\partial \varepsilon}{\partial T}}
\right) - \frac{1}{2pT} K \hat{F}^{<ki>}q_i \, ,
\end{eqnarray}
which is exactly the same which can be obtained from the calculations in \cite{1} and in
the particular case with all the symmetry conditions, that is $L=\frac{5}{6} K$.

\subsection{A comparison with the results of \cite{1}.}

We have already seen that the entropy density found with the present approach has at
equilibrium the same expression than in \cite{1}; similarly, for the first order
expression $h^{(1)}=0$ and for the second order expression (\ref{ag.1}). Similarly, for
the non convective part of the entropy flux in both articles it is zero at equilibrium,
has the form (\ref{ah.2}) at first order with respect to equilibrium, and the same form
(\ref{ai.1}) at second order.  \\
Regarding the non convective part of $F^{kij}$, in both papers it is zero at equilibrium
and have the same expression at first order, that is (\ref{ai.2}) of the present paper
and $(47)_{1,2}$ of \cite{1}. \\
Regarding $G^{kill}$, its expression at equilibrium here is given by $(\ref{cris14})_2$
which is the same of \cite{1}, except for identifying the present function $\phi_{0,0,1}$
with the function $\beta_1$ of \cite{1}, as it has be done in eq. (\ref{cris26}). \\
For what concerns $G^{kill}_{(1)}$, we obtain the same expression if we link the function
$\phi_{0,1,1}$ of the present paper with the function $K$ of \cite{1} through eq.
$(\ref{af.3})_2$; the expressions are $(\ref{af.3})_1$ in the present paper and
$(47)_{3}$ of \cite{1}, obviously in the case $L=\frac{5}{6} K$ of full symmetries. \\
It  is worth noting that in $(48)_{2}$ of \cite{1} we find also the definition of
$\beta_3$ which can be also interpreted as the definition of $K$, that is
\begin{eqnarray}\label{ai.3}
&{}& K=   \frac{1}{h_4} \left[ \beta_3 - 4 h_3 \left(\frac{p}{\rho}+ \varepsilon \right)
\right] \, .
\end{eqnarray}
Moreover, in $(48)_{1}$ of \cite{1} also $\beta_2$ is defined and under the further
assumption of full symmetries (that is $L=\frac{5}{6} K$), this definition is equivalent
to
\begin{eqnarray}\label{al.1}
&{}& \beta_2 = \left( 4 h_2 - \frac{10}{3} h_3 \right)  \left(\frac{p}{\rho}+ \varepsilon
\right) + \frac{5}{6} \, \beta_3 \, .
\end{eqnarray}
Thanks to this result, eqs. $(44)_{2,5}$ of \cite{1} become consequences of $(44)_{3,6}$
of \cite{1}. (Note that the integrability condition on $(44)_{1,4}$ allows to deduce
$\frac{\partial h_4}{\partial \rho}$).  \\
By substituting (\ref{ai.3}) in (\ref{af.3}) we find
\begin{eqnarray}\label{al.2}
&{}& \phi_{0,0,1} = \frac{3}{2} \, \beta_3
\end{eqnarray}
which links the function $\phi_{0,0,1}$ of the present article with the function
$\beta_3$ of \cite{1}. After that,
\begin{itemize}
  \item The equations $(43)_{1,2}$ of \cite{1} are just eqs. (\ref{cris31}) and (\ref{ag.2})
  of the present article,
  \item eqs. $(44)_{1,4}$ of \cite{1} are the same of the present eqs. (\ref{cris25}) and (\ref{cris27}),
  \item eqs. $(44)_{3,6}$ of \cite{1} (with $\beta_2$ given by the above (\ref{al.1}) ) coincide with the
  present eqs. (\ref{z.2}) and (\ref{alfa.5}).
\end{itemize}

\appendix

\section{Appendix 1: The particular solution $H=H_1$.}
Let us prove that $H=H_1$, with $H_1$ given by (\ref{11.6}) and $\psi_n$ constrained by
(\ref{11.5}), is a particular solution of (\ref{9.1}) and (\ref{9.3}). \\
In fact, it is easy to see that it satisfies $(\ref{9.1})_{3,4}$. \\
By substituting (\ref{11.6}) in $(\ref{9.1})_1$, we obtain \\
\begin{eqnarray*}
\frac{\partial^{r+p+1}}{\partial \lambda^r
\partial \mu^{p+1}} \left[ \left( \frac{-1}{2 \lambda} \right)^{q+1+\frac{p+r}{2}}
\psi_{\frac{p+r}{2}}\right] = \frac{\partial^{r+p+2}}{\partial \lambda^r
\partial \mu^{p+2}} \left[ \left( \frac{-1}{2 \lambda} \right)^{q+\frac{p+2+r}{2}}
\psi_{\frac{p+2+r}{2}}\right]
\end{eqnarray*}
which surely holds because $\psi_{\frac{p+r}{2}}= \frac{\partial}{\partial \mu}
\psi_{\frac{p+r}{2}+1}$, thanks to (\ref{11.5}).  \\
By substituting (\ref{11.6}) in $(\ref{9.1})_2$, we obtain \\
\begin{eqnarray*}
\frac{\partial^{r+p+2}}{\partial \lambda^{r+1}
\partial \mu^{p+1}} \left[ \left( \frac{-1}{2 \lambda} \right)^{q+\frac{p+r+1}{2}}
\psi_{\frac{p+r+1}{2}}\right] =\frac{\partial^{r+p+2}}{\partial \lambda^{r+1}
\partial \mu^{p+1}} \left[ \left( \frac{-1}{2 \lambda} \right)^{q+\frac{p+1+r}{2}}
\psi_{\frac{p+1+r+}{2}}\right]
\end{eqnarray*}
which is an evident identity. \\
It is more delicate to verify (\ref{9.3}). To do it, let us substitute (\ref{9.3}) with
its derivatives with respect to $\mu_{i_1}$, $\cdots$ , $\mu_{i_P}$, $\mu_{h_1k_1}$,
$\cdots$ , $\mu_{h_Qk_Q}$, $\lambda_{j_1}$, $\cdots$ , $\lambda_{j_R}$; let us substitute
(\ref{11.6}) in the resulting equation and let us calculate the last form at equilibrium.
We obtain
\begin{eqnarray*}
0 = P \delta^{i \overline{i_1}} \delta^{(\overline{i_2 \cdots i_{P}}kh_1k_1 \cdots
h_Qk_Qj_1 \cdots j_R)} \frac{(P+2Q+R+1)!!}{P+2Q+R+1} \frac{\partial^{R+P+1}}{\partial
\lambda^R
\partial \mu^{P+1}} \left[ \left( \frac{-1}{2 \lambda} \right)^{Q+\frac{P+R}{2}}
\psi_{\frac{P+R}{2}}\right] + \\
+2 Q  \delta^{i \overline{h_1}} \delta^{(\overline{k_1 h_2k_2 \cdots h_Qk_Q}ki_1 \cdots
i_{P}j_1 \cdots j_R)} \frac{(P+2Q+R+1)!!}{P+2Q+R+1} \frac{\partial^{R+P+1}}{\partial
\lambda^R
\partial \mu^{P+1}} \left[ \left( \frac{-1}{2 \lambda} \right)^{Q+\frac{P+R}{2}}
\psi_{\frac{P+R}{2}}\right] + \\
+ 2 \lambda   \delta^{(ki h_1k_1 \cdots h_Qk_Q i_{1} \cdots i_{P}j_1 \cdots j_R)}
(P+2Q+R+1)!! \frac{\partial^{R+P+1}}{\partial \lambda^R
\partial \mu^{P+1}} \left[ \left( \frac{-1}{2 \lambda} \right)^{Q+1+\frac{P+R}{2}}
\psi_{\frac{P+R}{2}}\right] + \\
+ 2 R   \delta^{(ki h_1k_1 \cdots h_Qk_Q i_{1} \cdots i_{P}j_1 \cdots j_R)} (P+2Q+R+1)!!
\frac{\partial^{R+P}}{\partial \lambda^{R-1}
\partial \mu^{P+1}} \left[ \left( \frac{-1}{2 \lambda} \right)^{Q+1+\frac{P+R}{2}}
\psi_{\frac{P+R}{2}}\right] + \\
+R \delta^{i \overline{j_1}} \delta^{(\overline{j_2 \cdots j_{R}}kh_1k_1 \cdots
h_Qk_Qi_{1} \cdots i_{P})} \frac{(P+2Q+R+1)!!}{P+2Q+R+1} \frac{\partial^{R+P+1}}{\partial
\lambda^R
\partial \mu^{P+1}} \left[ \left( \frac{-1}{2 \lambda} \right)^{Q+\frac{P+R}{2}}
\psi_{\frac{P+R}{2}}\right] + \\
+\delta^{ki} \delta^{(i_{1} \cdots i_{P} h_1k_1 \cdots h_Qk_Q j_1 \cdots j_R)}
\frac{(P+2Q+R+1)!!}{P+2Q+R+1} \frac{\partial^{R+P+1}}{\partial \lambda^R
\partial \mu^{P+1}} \left[ \left( \frac{-1}{2 \lambda} \right)^{Q+\frac{P+R}{2}}
\psi_{\frac{P+R}{2}}\right] \, ,
\end{eqnarray*}
where overlined indexes denote symmetrization over those indexes, after that the other
one (round brackets around indexes) has been taken. \\
Now, the first, second, fifth and sixth term can be put together so that the above
expression becomes
\begin{eqnarray*}
0 = \delta^{i \overline{i_1}} \delta^{(\overline{i_2 \cdots i_{P}kh_1k_1 \cdots h_Qk_Qj_1
\cdots j_R})} (P+2Q+R+1)!! \frac{\partial^{R+P+1}}{\partial \lambda^R
\partial \mu^{P+1}} \left[ \left( \frac{-1}{2 \lambda} \right)^{Q+\frac{P+R}{2}}
\psi_{\frac{P+R}{2}}\right] + \\
+ (P+2Q+R+1)!! \delta^{(ki h_1k_1 \cdots h_Qk_Q i_{1} \cdots i_{P}j_1 \cdots j_R)}
\left\{ 2 \lambda    \frac{\partial^{R+P+1}}{\partial \lambda^R
\partial \mu^{P+1}} \left[ \left( \frac{-1}{2 \lambda} \right)^{Q+1+\frac{P+R}{2}}
\psi_{\frac{P+R}{2}}\right] \right. + \\
+ \left. 2 R \frac{\partial^{R+P}}{\partial \lambda^{R-1}
\partial \mu^{P+1}} \left[ \left( \frac{-1}{2 \lambda} \right)^{Q+1+\frac{P+R}{2}}
\psi_{\frac{P+R}{2}}\right] \right\} \, ,
\end{eqnarray*}
which is satisfied as a consequence of the property \\
$\delta^{i \overline{i_1}} \delta^{(\overline{i_2 \cdots i_{P}kh_1k_1 \cdots h_Qk_Qj_1
\cdots j_R})} = \delta^{(ki h_1k_1 \cdots h_Qk_Q i_{1} \cdots i_{P}j_1 \cdots j_R)}$ and
of the identity
\begin{eqnarray*}
\frac{\partial^{R}}{\partial \lambda^R} \left[ \left( \frac{-1}{2 \lambda}
\right)^{Q+\frac{P+R}{2}} \psi_{\frac{P+R}{2}}\right] = \frac{\partial^{R}}{\partial
\lambda^R} \left[ - 2 \lambda \left( \frac{-1}{2 \lambda} \right)^{Q+1+ \frac{P+R}{2}}
\psi_{\frac{P+R}{2}}\right]  = \\
= - 2 \lambda \frac{\partial^{R}}{\partial \lambda^R} \left[ \left( \frac{-1}{2 \lambda}
\right)^{Q+1+ \frac{P+R}{2}} \psi_{\frac{P+R}{2}}\right] -2R
\frac{\partial^{R-1}}{\partial \lambda^{R-1}} \left[ \left( \frac{-1}{2 \lambda}
\right)^{Q+1+\frac{P+R}{2}} \psi_{\frac{P+R}{2}}\right] \, .
\end{eqnarray*}
This completes the  proof that $H=H_1$ is a particular solution of (\ref{9.1}) and
(\ref{9.3}).

\section{Appendix 2: Proof of the PROPERTY 1.}

Let us the property 1 with the iterative procedure and let $\Delta H^n$ denote the
homogeneous part of $\Delta H$ of order $n$ with respect to equilibrium. We have, \\
\begin{itemize}
  \item Case $n=1$: The equation $(\ref{11.9})_5$ at equilibrium, thanks to (\ref{11.10})
  becomes $2 \frac{\partial^2 \Delta H^1}{\partial \mu \partial  \mu_{ki}} \lambda =0$ from which we have that
  $\frac{\partial \Delta H^1}{\partial \mu }$ can depend only on $\mu$, $\mu_i$,
  $\lambda$, $\lambda_c$; but the representation theorems show that no scalar function of
  order $1$ with respect to equilibrium can depend only on these variables. It follows
  that $\frac{\partial \Delta H^1}{\partial \mu }=0$ so that $\Delta H^1$ is of degree
  zero with respect to $\mu$ and the property is verified for this case.
 \item Case $n \geq 2$: Let us suppose, for the iterative hypothesis that $\Delta H$ up to order $n \geq 1$ with respect to
equilibrium is a polynomial of degree $n-1$ in the variable $\mu$; we proceed now to
prove that this property holds also with $n+1$ instead of $n$.
\end{itemize}
In fact, eq. $(\ref{11.9})_1$ up to order $n-1$ gives \\
$\frac{\partial^2 \Delta H^n}{\partial  \mu \partial  \mu_{ij}} = \frac{\partial^2 \Delta
H^{n+1}}{\partial \mu_i \partial  \mu_{j}}$ from which we have
\begin{eqnarray}\label{p1.1}
\Delta H^{n+1} = P_{n-2} + \Delta H^{n+1}_i( \mu, \mu_{ab}, \lambda, \lambda_c) \mu^i +
\Delta H^{n+1}_0( \mu, \mu_{ab}, \lambda, \lambda_c)
\end{eqnarray}
where $P_{n-2}$ is a polynomial of degree $n-2$ in  $\mu$ and which is at least quadratic
in $\mu_j$. \\
After that, eq. $(\ref{11.9})_2$ up to order $n$ gives \\
$\frac{\partial^2 \Delta H^{n+1}}{\partial \mu
\partial  \lambda_i} = \frac{\partial^2 \Delta H^{n+1}}{\partial \lambda \partial
\mu_{i}}$ \\
which, thanks to (\ref{p1.1}) becomes
\begin{eqnarray}\label{p2.1}
\frac{\partial^2  P_{n-2}}{\partial \mu
\partial  \lambda_i}  + \frac{\partial^2  \Delta H^{n+1}_j}{\partial \mu
\partial  \lambda_i} + \frac{\partial^2  \Delta H^{n+1}_0}{\partial \mu
\partial  \lambda_i} = \frac{\partial^2  P_{n-2}}{\partial \lambda \partial
\mu_{i}} + \frac{\partial  \Delta H^{n+1}_i}{\partial \lambda } \, .
\end{eqnarray}
This relation, calculated in $\mu_j=0$ gives
\begin{eqnarray}\label{p2.2}
\frac{\partial^2  \Delta H^{n+1}_0}{\partial \mu
\partial  \lambda_i} = \frac{\partial  \Delta H^{n+1}_i}{\partial \lambda }
\end{eqnarray}
because $P_{n-2}$ is at least quadratic in $\mu_j$. \\
The derivative of (\ref{p2.1}) with respect to $\mu_j$, calculated then in $\mu_j=0$, is
\\
$\frac{\partial^2  \Delta H^{n+1}_j}{\partial \mu
\partial  \lambda_i} = \left( \frac{\partial^3  P_{n-2}}{\partial \mu_{j} \partial \lambda
\partial \mu_{i}} \right)_{ \mu_{j}=0}$ from which $\frac{\partial  \Delta H^{n+1}_j}{
\partial  \lambda_i} = P^{ij}_{n-1}$ with $P^{ij}_{n-1}$ a polynomial of degree $n-1$ in
$\mu$. By integrating this relation, we obtain \\
$\Delta H^{n+1}_i= P^{i}_{n-1} + f^{i}_{n-1}(\mu, \mu_{ab}, \lambda,)$  \\
where $P^{i}_{n-1}$ is a polynomial of degree $n-1$ in $\mu$. But, for the Representation
Theorems, a vectorial function such as $f^{i}_{n-1}$ is zero because it depends only on
scalars and on a second order tensor. It follows that
\begin{eqnarray}\label{p3.1}
\Delta H^{n+1}_i= P^{i}_{n-1} \, .
\end{eqnarray}
By using this result, eq. (\ref{p2.2}) can be integrated and gives
\begin{eqnarray}\label{p3.2}
\frac{\partial  \Delta H^{n+1}_0}{ \partial  \lambda_i} = P^{i}_{n}
\end{eqnarray}
with $P^{i}_{n}$  a polynomial of degree $n$ in $\mu$. \\
Now we impose eq. $(\ref{11.9})_5$ at order $n$ and see that its first, second, fifth and
sixth terms are of degree $n-2$ in $\mu$ so that we have \\
$2 \frac{\partial^2 \Delta H^{n+1}}{\partial \mu
\partial  \mu_{ki}}\lambda +
2 \frac{\partial^2 \Delta H^{n+1}}{\partial  \mu_k \partial  \mu_{ij}} \lambda_{j} =
Q_{n-2}$  \\
with $Q_{n-2}$  a polynomial of degree $n-2$ in $\mu$. This relation, thanks to
(\ref{p1.1}) becomes \\
$2 \lambda \frac{\partial^2 \Delta H_a^{n+1}}{\partial \mu
\partial  \mu_{ki}} \mu^a + 2 \lambda \frac{\partial^2 \Delta H_0^{n+1}}{\partial \mu
\partial  \mu_{ki}} + 2 \lambda_{j} \frac{\partial^2 P_{n-2}}{\partial  \mu_k \partial
\mu_{ij}}+ 2 \lambda_{j} \frac{\partial }{\partial  \partial \mu_{ij}} \Delta H_k^{n+1} =
Z_{n-2}$ \\
with $Z_{n-2}$  a polynomial of degree $n-2$ in $\mu$. \\
This relation,  calculated in $\mu_j=0$, thanks to (\ref{p3.1}) and to the fact that
$P_{n-2}$ is at least quadratic in $\mu_j$, gives \\
$2 \lambda \frac{\partial^2 \Delta H_0^{n+1}}{\partial \mu
\partial  \mu_{ki}} = \bar{Q}_{n-1}^{ki}$ \\
with $\bar{Q}_{n-1}^{ki}$ a polynomial of degree $n-1$ in $\mu$. It follows that \\
$\frac{\partial \Delta H_0^{n+1}}{
\partial  \mu_{ki}} = \bar{P}_{n}^{ki}$ \\
with $\bar{P}_{n}^{ki}$ a polynomial of degree $n$ in $\mu$. This result, jointly with
(\ref{p3.2}) gives that
\begin{eqnarray}\label{p4.1}
\Delta H_0^{n+1} = \tilde{P}_{n} + f(\mu, \lambda) \, .
\end{eqnarray}
But a function depending only on $\mu$ and $\lambda$ cannot be of order $n+1$ with
respect to equilibrium; it follows that $f(\mu, \lambda)=0$. \\
Consequently, (\ref{p4.1}), (\ref{p3.1}) and (\ref{p1.1}) give that $\Delta H^{n+1}$ is a
polynomial of degree $n$ in $\mu$ and this completes the proof. \\
We note also that the proof of PROPERTY 1 has not required the conditions
$(\ref{11.9})_{3,4}$ so that it holds also without the symmetry conditions
$(\ref{3.0})_{4,5}$; this will be useful for a future work when we will study this model
without these two last symmetry conditions.

\end{document}